\documentclass[11pt]{article}
\usepackage{amsmath,amssymb,color,epsfig,cite}
\usepackage{xcolor}
\usepackage{mathrsfs}
\usepackage{float}

\usepackage{mathtools}

\usepackage{amsfonts}

\newcommand{\hoch}[1]{$\, ^{#1}$}

\makeatletter
\@addtoreset{equation}{section}
\makeatother

\newcommand{\be}{\begin{equation}}
\newcommand{\ee}{\end{equation}}
\newcommand{\bea} {\begin{eqnarray}}
\newcommand{\eea}{\end{eqnarray}}
\newcommand{\nn}{\nonumber}

\def\ft#1#2{{\textstyle{\frac{\scriptstyle #1}{\scriptstyle #2} } }}
\def\fft#1#2{{\frac{#1}{#2}}}

\def\0{{\sst{(0)}}}
\def\1{{\sst{(1)}}}
\def\2{{\sst{(2)}}}
\def\3{{\sst{(3)}}}
\def\4{{\sst{(4)}}}
\def\5{{\sst{(5)}}}
\def\6{{\sst{(6)}}}
\def\7{{\sst{(7)}}}
\def\8{{\sst{(8)}}}
\def\sst#1{{\scriptscriptstyle #1}}

\def\ep{{\epsilon}}
\def\del{{\partial}}

\def\cF{{{\cal F}}}

\def\R{{\mathbb R}}

\def\ben{\begin{equation}}
\def\bea{\begin{eqnarray}}
\def\een{\end{equation}}
\def\eea{\end{eqnarray}}

\def\ft#1#2{{\textstyle{\frac{\scriptstyle #1}{\scriptstyle #2} } }}
\def\fft#1#2{{\frac{#1}{#2}}}

\begin{document}

\begin{flushright}
\hfill {UPR-1321-T\ \ \ MI-HET-798}\\
\end{flushright}

\begin{center}

{\large {\bf Mass And Force Relations For 
Einstein-Maxwell-Dilaton
Black Holes}}

\vspace{15pt}
{\large S. Cremonini$^{1,2}$, M. Cveti\v c$^{3,4,5}$, 
              C.N. Pope$^{6,7}$ and A. Saha$^{6}$}

\vspace{15pt}

{\hoch{1}}{\it Department of Physics, Lehigh University, Bethlehem, 
PA 18018, USA}

{\hoch{2}}{\it 
Kavli Institute of Theoretical Physics, University of California Santa Barbara, Santa Barbara,CA, 93106, USA}

{\hoch{3}}{\it Department of Physics and Astronomy,
University of Pennsylvania, \\
Philadelphia, PA 19104, USA}

{\hoch{4}}{\it Department of Mathematics, University of Pennsylvania, Philadelphia, PA 19104, USA}

{\hoch{5}}{\it Center for Applied Mathematics and Theoretical Physics,\\
University of Maribor, SI2000 Maribor, Slovenia}


\hoch{6}{\it George P. \& Cynthia Woods Mitchell  Institute
for Fundamental Physics and Astronomy,\\
Texas A\&M University, College Station, TX 77843, USA}

\hoch{7}{\it DAMTP, Centre for Mathematical Sciences,
 Cambridge University,\\  Wilberforce Road, Cambridge CB3 OWA, UK}

\vspace{10pt}

\end{center}

\begin{abstract}

We investigate various properties of extremal dyonic static black holes in 
Einstein-Maxwell-Dilaton theory.
Using the fact that 
the long-range force 
between two identical extremal black holes always vanishes,  we obtain a simple first-order
ordinary differential equation for 
the black hole mass in terms of its electric and
magnetic charges.  
Although this
equation appears not to be solvable explicitly for general values of the
strength $a$ of the dilatonic coupling to the Maxwell field, it 
nevertheless provides a powerful way of characterising the black hole mass
and the scalar charge. 
We make use of these
expressions to derive general results about the long-range force
 between two non-identical extremal black holes.  In particular,
we argue that the force is repulsive whenever $a>1$ and attractive 
whenever $a<1$ (it vanishes in the intermediate BPS case $a=1$). 
The sign of the force is also correlated with the 
sign of the binding energy between extremal black holes, as well as with the convexity or concavity of the surface characterizing the extremal mass as a function of the charges.
Our work is motivated in part by the Repulsive Force Conjecture and the question of whether long range forces between non-identical states can shed new light on the Swampland.

\end{abstract}

{\scriptsize 
cremonini@lehigh.edu, cvetic@physics.upenn.edu,
pope@physics.tamu.edu, aritrasaha@physics.tamu.edu.}

\pagebreak

\tableofcontents
\addtocontents{toc}{\protect\setcounter{tocdepth}{2}}

\newpage

\section{Introduction}

Recent years have seen growing efforts to sharpen the constraints that theories of quantum gravity place on  
low energy effective field theories (EFTs). 
Within these efforts, one of the challenges has been to quantify  
the notion that 
gravity is the weakest force 
and to understand what it tells us about the structure of long range interactions, 
in theories with a quantum gravity UV completion.
In particular, the attempts to understand this question have led to various generalizations of the Weak Gravity Conjecture (WGC) 
 \cite{Arkani-Hamed:2006emk} 
and extensive studies of its phenomenological consequences (see \emph{e.g.} \cite{Harlow:2022gzl} for a comprehensive review).
In its simplest form, the WGC requires the existence of superextremal charged particles, states whose mass is smaller than or equal to their charge (in Planck units).
In flat space, a closely related -- but not equivalent\footnote{In theories of quantum gravity with massless scalars, the WGC and the RFC are distinct. The differences become even more apparent when one takes into account the effects of higher derivative corrections \cite{Cremonini:2021upd,Etheredge:2022rfl}.} -- way to quantify the weakness of gravity has led to the  Repulsive Force Conjecture (RFC) 
\cite{Arkani-Hamed:2006emk,Palti:2017elp,Heidenreich:2019zkl},
 which roughly states that theories compatible with quantum gravity should contain \emph{self-repulsive states}, 
 i.e. states which would feel either a repulsive or vanishing force when placed asymptotically far from an identical copy of themselves. 
 Both the WGC and the RFC place restrictions on low energy EFTs
 and have implications for the spectrum of states in the theory.
 Moreover, the RFC constrains all the interactions that lead to long range self-forces, 
 including those coming from massless scalar fields.
 A question that arises naturally, then, is what these conjectures 
 teach us about binding energies and the existence of bound states.

Thus far most of the discussions of the RFC have centered around studies of 
long range interactions between two copies of the \emph{same} state, i.e. self-forces.
However, it may be useful to explore what happens to the force and binding energy 
when the states in consideration are \emph{not} the same, using black holes as probes.
Indeed, while asymptotically the force between two identical extremal black holes is known \cite{Heidenreich:2020upe}
to be zero,\footnote{\label{identfoot} 
In this context, ``identical'' can mean that
the set of electric and magnetic charges carried by the two extremal black
holes are either exactly the same, or, more generally, 
that the set of charges carried
by one of the black holes is an overall positive constant multiple of the
set of charges carried by the other.} 
 even in the presence of scalar matter, 
very little is known
about its structure when the black holes carry different charges and represent distinct states. 
Using non identical states as probes of long range interactions may uncover novel features 
and potentially new insights on the string theory Landscape. 
As a concrete example, in \cite{Cremonini:2022sxf} we saw that the long-range force between distinct, extremal KK dyonic black holes is \emph{always repulsive}. While this is naively surprising, it might have a natural explanation, 
at the microscopic level, in terms of the interactions between the constituent D-branes (in this case D0-D6 branes) 
and properties of bound states in the theory.  
If this is indeed the case, long range forces might provide an easier way to access some of the information encoded in the microscopic description of the theory.

Motivated by the questions above and by our previous work \cite{Cremonini:2022sxf}, 
in this paper our goal is to understand
whether long range forces between different extremal black holes display 
any \emph{generic} features, and, if so, how the latter are correlated with specific properties of the theory they arise in. 
As we will see, the scalar couplings in the theory we examine will leave clear imprints on certain characteristics of the black hole solutions (such as their extremality relations and propensity to bind), 
which will then be imprinted on the behavior of the long range interactions between them.

 We are going to work with extremal static black holes solutions to  four-dimensional 
Einstein-Maxwell-Dilaton theory, described by the Lagrangian
\bea
{\cal L}= \sqrt{-g}\, \big( R -\ft12(\del\phi)^2 - \ft14 e^{a\phi}\, 
  F^{\mu\nu} F_{\mu\nu}\big)\,,\label{emdlag}
\eea
where the constant $a$ characterises the strength of the exponential coupling
of the dilaton to the Maxwell field.  For certain values of $a$, namely
$a=0$, $a=1$ and $a=\sqrt3$, the solutions for dyonic black holes, carrying
both electric and magnetic charge, are known explicitly.  
   Our focus will be on the properties of the extremal static dyonic black hole
solutions for \emph{arbitrary} values of $a$.\footnote{\label{anfoot} 
Some properties of
these extremal black holes were investigated numerically in \cite{poltwawil}
and analytically in \cite{galkhrorl}.  The solutions exist for all values
of $a$, but there is some degree of non-analyticity on the horizon unless
$a$ is such that $a^2=\ft12 k(k+1)$ where $k$ is an integer.  Scalar
curvature invariants, such as $R^{\mu\nu\rho\sigma}\, R_{\mu\nu\rho\sigma}$,
$\nabla^\lambda R^{\mu\nu\rho\sigma}\,\nabla_\lambda R_{\mu\nu\rho\sigma}$,
etc., are finite on the horizon, however, for all values of $a$.}  
The mass $M$ of such an extremal 
black hole
will be a function of the electric and magnetic charges $Q$ and $P$,
with $M=\cF(Q,P)$,
but except for the exactly-solvable cases when $a=0$, 1 or $\sqrt3$,
the explicit form of the function $\cF(Q,P)$ is unknown. One of the main
results in our work is 
a simple nonlinear 
first-order ordinary differential
equation for 
(a rescaled version of) $\cF(Q,P)$.
Although this equation 
is, as far as we know, exactly solvable only at the special values
$a=1$ and $\sqrt3$, the fact that we can express 
the mass 
in this 
relatively simple way allows us to probe a number of properties of the
extremal black holes.  

 The long-range force
$F_{12}$ between two black holes takes the form
\bea
F_{12}= \fft1{r^2}\, \Big[ Q_1\, Q_2 + P_1\, P_2 - \ft14 M_1\, M_2 -
\Sigma_1\,\Sigma_2\Big]\,,\label{F120}
\eea
where $Q$, $P$, $M$ and $\Sigma$ denote, respectively, the electric and magnetic
 charges\footnote{For all the non-identical extremal black holes that we
shall consider, we shall always take the {\it signs} of the charges
of the first and the second black hole to be the same.
The case corresponding to opposite signs for the 
charges would be of little interest, since the electrostatic forces would just reinforce the gravitational attraction.  
Thus, without loss of generality, we assume that all the electric and
magnetic charges are positive.},
the mass and the scalar charge, while the subscripts $1,2$ are used to 
distinguish the first from the second black hole.  
In our earlier work \cite{Cremonini:2022sxf} we initiated a study of (\ref{F120}), 
focusing on a specific class of Toda theories which support extremal black holes that are not BPS.
Here we extend our analysis to a much broader class of dyonic solutions, and 
identify a number of new features.
Most notably, the range of the parameter $a$ controlling the gauge kinetic coupling dictates certain geometric 
properties of the energy surface $M=\cF(Q,P)$ of each extremal solution, as well as the sign of the long range interactions between distinct ones.

In particular, using a combination of
approximations, and also numerical analysis, we conclude that the force
between non-identical extremal dyonic black holes is always repulsive if the
dilaton coupling $a$ appearing in (\ref{emdlag}) satisfies $a>1$, and it is always attractive if
$a<1$.  In the intermediate case $a=1$, for which the extremal dyonic 
black holes are BPS, the force between them is always zero.
Moreover, using geometrical arguments we show that when $a>1$ the energy surface $M=\cF(Q,P)$ describing the mass of each solution is \emph{convex}, while 
for $a<1$ it is \emph{concave}.

We also investigate the related question of what is the sign of the binding
energy between extremal black holes. Thus, if the mass of an extremal black hole
with charges $Q$ and $P$ is $M=\cF(Q,P)$, we define the binding
energy between two such black holes to be
\bea
\Delta M= \cF(Q_1+Q_2,P_1+P_2) - \cF(Q_1,P_1) -\cF(Q_2,P_2)\,.
\eea
Intuitively, one may expect that if $\Delta M$ is positive then the
two constituent black holes with charges $(Q_1,P_1)$ and $(Q_2,P_2)$
should tend to repel one another, while if $\Delta M$ is negative they
should attract.  
Indeed, we find that the sign of 
$\Delta M$ does correlate with the sign of the long-range force, for
all the EMD extremal black holes. Perhaps not surprisingly, the sign of $\Delta M$ is also governed by the convexity
or concavity of the surface $M=\cF(Q,P)$.

The paper is organised as follows.  
In section 2 we describe properties of the extremal EMD black hole solutions we  will be working with.
In section 3 we derive a simple differential equation that controls the mass of the extremal black hole in terms of its electric and magnetic charges, and in section 4 we present some solutions valid in specific perturbative regimes.
Section 5 is devoted to the computation of the long distance force between non-identical black holes, while section 6 discusses the binding energies. 
  Geometrical properties of the energy surface describing how the mass is related to the charges are discussed in section 7.
  Finally, in Appendix A we include some examples 
of the numerical computations that support our results, while in Appendix B we prove certain properties of the binding energy near the special value of the coupling $a=1$.

\section{Static Extremal Black Holes in Einstein-Maxwell-Dilaton Theory}

  Purely electric or magnetic static black holes in the EMD theory (\ref{emdlag}) for 
arbitrary values of the dilaton coupling were constructed in 
\cite{gibbmaed}.  The system of equations for the most general
dyonic static solutions was obtained in \cite{lupoxu}, where it was shown
that they could be reduced to a Toda-like system.  It was noted there that
the equations became exactly those of the $SU(3)$ Toda system when
$a=\sqrt3$, and of the $SU(2)\times SU(2)$ Toda (or (Liouville)$^2$) system
when $a=1$, but that no explicit dyonic solutions could be obtained for
generic values of the dilaton coupling.  The dyonic solution for
$a=\sqrt3$ had been obtained by \cite{gibbwilt,gibbkall}.  

  A formulation
of the Toda-like equations for general values of $a$ appeared also in a
recent paper \cite{lustea}.  With some adaption of their notation to suit
our conventions, the static  
black hole solutions to the equation of motion 
following from the Lagrangian (\ref{emdlag}) are given by
\bea
ds^2 &=& - f(r)\, dt^2 + \fft{dr^2}{f(r)} + r^2\, (H_e H_m)^h\,
  d\Omega_2^2\,,\nn\\
f(r)&=& (H_e H_m)^{-h}\, \Big(1-\fft{\mu}{r}\Big)\,,\nn\\
e^{a\phi} &=& \Big(\fft{H_e}{H_m}\Big)^{2-h}\,,\nn\\
F &=& \fft{Q}{r^2}\, H_e^{-2}\, H_m^{2-2h}\, dt\wedge dr + 
   P\, \sin\theta d\theta\wedge d\varphi\,,\label{solans}
\eea
where $Q$ and $P$ denote the electric and magnetic charges, the parameter $\mu$ is positive for non-extremal black holes and
equal to zero for extremal black holes, and 
\bea
h= \fft{2}{1+a^2}\,.
\eea
Defining the inverse radial coordinate 
\bea
\rho= \fft1{r}\,,
\eea
 the
EMD equations of motion imply that 
the functions $H_e$ and $H_m$ obey the equations
\bea
\Big(\fft{ (1-\mu\rho)\, H_e'}{H_e}\Big)' +
\fft{8Q^2\, H_m^{2-2h}}{h\, H_e^2}=0\,,\qquad
\Big(\fft{ (1-\mu\rho)\, H_m'}{H_m}\Big)' +
\fft{8 P^2\, H_e^{2-2h}}{h\, H_m^2}=0\,,\label{HeHmeom}
\eea
where a prime denotes a derivative with respect to $\rho$.
Since our focus in this paper is on extremal black holes
we shall set $\mu=0$ from now on. The EMD equations of motion also imply 
a constraint, which for $\mu=0$ reads 
\bea
\fft{{H_e'}^2}{H_e^2} + \fft{{H_m'}^2}{H_m^2} +
  \fft{2(h-1)\, H_e' H_m'}{H_e H_m} - \fft{8 Q^2\, H_m^{2-2h}}{h\, H_e^2}
  -\fft{8 P^2\, H_e^{2-2h}}{h\, H_m^2}=0\,.\label{con1}
\eea

  As already mentioned, except for the cases $a=1$ and $a=\sqrt3$ (and, more
trivially, $a=0$), for which the equations are exactly solvable, no
explicit solutions for dyonic black holes are known.  One approach to
studying the solutions in general is to look first for solutions as 
series expansions in the asymptotic region where $r$ goes to infinity, 
which corresponds to $\rho\longrightarrow 0$.  Thus one
may seek solutions of the form
\bea
H_e(\rho)= 1 + e_1\, \rho + e_2\,\rho^2 + e_3\,\rho^3 + \cdots\,,\qquad
H_m(\rho)= 1 + m_1\, \rho + m_2\,\rho^2 + m_3\,\rho^3 + \cdots\,,
\label{HeHmexp}
\eea
where the $e_i$ and $m_i$ are constants.  Substituting these expansions into (\ref{HeHmeom})
with $\mu=0$ (i.e., working at extremality), one finds that all the $e_i$ and $m_i$ for $i\ge 3$ can be
solved for in terms of $(e_1,e_2,m_1,m_2)$, with the electric and 
magnetic charges $Q$ and $P$ being given by
\bea
Q^2= 2h\, (e_1^2 -2 e_2)\,,\qquad P^2 = 2h\, (m_1^2 - 2 m_2)\,.
\label{QPeimi}
\eea
The constraint equation (\ref{con1}) implies one condition on the
four free parameters $(e_1,e_2,m_1,m_2)$, namely
\bea
e_2 + m_2 + (h-1) e_1\, m_1=0\,.\label{con2}
\eea
We shall view the constraint as determining $m_2$ in terms of the three
remaining parameters $(e_1,e_2,m_1)$.

   We know that the extremal dyonic black hole solutions should be 
characterised by just two parameters and not three.  The
three-parameter solutions will in fact generically 
describe \emph{singular} spacetimes.  This can be seen by performing a numerical
integration of the equations, using the small-$\rho$ (i.e. large-$r$)
expansions characterised by $(e_1,e_2,m_1)$ to set initial data for
an integration to large $\rho$ (i.e. approaching the horizon at $r=0$,
which is $\rho=\infty$).  For generic choices of $(e_1,e_2,m_1)$ the solution
develops singularities at some finite value of $\rho$.  By choosing values
for two of the parameters (say $e_1$ and $m_1$), and then fine-tuning
$e_2$, one can home in on a single specific value for $e_2$ that gives a 
solution that behaves properly for a black hole spacetime.  In fact, 
the proper behaviour is such that $H_e(\rho)$ and $H_m(\rho)$ go like
\bea
H_e(\rho)\sim c_e\, \rho^{1/h}\,,\qquad H_m\sim c_m\, \rho^{1/h}
\label{nearhorizon}
\eea
at large $\rho$, so that the horizon has a finite and nonzero area 
(see the expression for the metric in eqns (\ref{solans})).  By means of
a ``shooting method'' approach, one can find the 
value of $e_2$, for given choices of $e_1$ and $m_1$ (and of course, also
of the dilaton coupling $a$) that gives the well-behaved black hole solution.
Since the mass of the extremal black hole is given by
\bea
M= 2 h\, (e_1 + m_1)\,,\label{Meimi}
\eea
this means that one can, very laboriously, collect numerical data that
would allow one to plot the mass as a function of the electric and 
magnetic charges.  (Recall that $Q$ and $P$ are specified in eqns (\ref{QPeimi}) in terms of the
$e_i$ and $m_i$ parameters).

In order to take into account the contribution of the massless scalar  in (\ref{emdlag}) to the long range force, we need to 
extract 
the scalar charge $\Sigma$, which can be defined as the coefficient of $r^{-1}$
in the large-$r$ expansion of the dilaton $\phi$,\footnote{The 
constant $\phi_\infty$, the asymptotic 
value of $\phi$ at infinity, is zero in the present
discussion.  However, it is sometimes useful to allow $\phi_\infty$ to
be non-zero, which can be accomplished using a global 
scaling symmetry of the EMD theory; see section \ref{MassDEQ} where we
discuss this and employ it in order to rewrite $\Sigma$ in terms of
derivatives of the mass with respect to the charges.}
\bea
\phi= \phi_\infty + \fft{\Sigma}{r} + {\cal O}(r^{-2})\,.\label{phiexp}
\eea
Thus one has
\bea
\Sigma= \sqrt{h(2-h)} \, (m_1-e_1)\,.\label{Sigeimi}
\eea
In principle, therefore, having accumulated sufficient numerical data
from computations that give the charges $Q$ and $P$, the mass $M$ and
the scalar charge $\Sigma$ for pairs of extremal black holes for a given 
choice of the dilaton coupling $a$, one could then calculate the force between
the distantly-separated black holes using eqn (\ref{F120}).  It is evident
that this would be a very time-consuming way of trying to learn about the
nature of the force as a function of the charges $Q$ and $P$ for different 
choices of the dilaton coupling.

  Note that by substituting the near-horizon behaviour of the functions
$H_e$ and $H_m$, as given in eqn (\ref{nearhorizon}), into the
equations of motion (\ref{HeHmeom}), we find that the constants
$c_e$ and $c_m$ are given by
\bea
c_e= 2^{\ft{3}{2h}}_{\phantom{Sigma\!\!\!\!\!\!\!\!}} 
Q^{-\ft1{h(h-2)}}\, P^{\ft{h-1}{h(h-2)}}\,,\qquad
c_m= 2^{\ft{3}{2h}}_{\phantom{Sigma\!\!\!\!\!\!\!\!}}\, 
P^{-\ft1{h(h-2)}}\, Q^{\ft{h-1}{h(h-2)}}\,.
\eea
From this, it follows that the area of the horizon is given by
\bea
{\cal A}_H = 32\pi\, Q\, P\,,
\eea
and so the entropy is $S=8\pi Q\, P$.
From these expressions it follows that purely electric or purely magnetic black holes have vanishing horizon area, i.e. they are ``small black holes.''

  One observation that is worth noting at this point is that if one
uses (\ref{F120}) to calculate the force between a pair of \emph{identical}
objects characterised by the four parameters $(e_1,m_1,e_2,m_2)$,
taking the charges, mass and scalar charge to be given by (\ref{QPeimi}),
(\ref{Meimi}) and (\ref{Sigeimi}), then one finds that the force
vanishes purely as a consequence of the constraint (\ref{con2}).  In other
words, regardless of whether one imposes the much more stringent 
condition that the solution should describe a genuine black hole, one
already finds just from the large-$r$ behaviour of the fields that
the force between two such identical objects will vanish.   The vanishing of
the force between two identical 
extremal black holes then follows as a consequence.  This latter result is
a particular manifestation of a general argument given in 
\cite{Heidenreich:2020upe}.
 
  We shall make use of the fact that the force between identical 
extremal black holes vanishes in the next section, when we derive a
simple equation that governs the mass of an extremal EMD black hole
in terms of its electric and magnetic charges.

\section{A Differential Equation For The Mass}
\label{MassDEQ}

  In the EMD theory, for general values of $a$, we know that the mass of
an extremal black hole carrying electric charge $Q$ and magnetic charge
$P$  
must be given by a formula of the form
\bea
M= \cF(Q,P)\,.\label{MfromcF}
\eea
The function $\cF(Q,P)$ will be different for different values
of the dilaton coupling $a$, and furthermore it is only known explicitly
for the three special cases $a=0$, $a=1$ and $a=\sqrt3$:
\bea
a=0:&& M= 2\sqrt{Q^2+P^2}\,,\nn\\
a=1:&& M= \sqrt2\, (Q+P)\,,\nn\\
a=\sqrt3:&& M=(Q^{\ft23} + P^{\ft23})^{\ft32}\,.\label{M3cases}
\eea
We do know
however, on dimensional grounds, 
that for any value of $a$ the mass function  
$\cF(Q,P)$ must be a homogeneous function of degree
1.  That is, we know that
\bea
\cF(\lambda Q,\lambda P)=\lambda\, \cF(Q,P)\,.
\eea
By differentiating this equation with respect to $\lambda$, and then
setting $\lambda=1$, it follows that
\bea
Q\fft{\del \cF(Q,P)}{\del Q} + P\, \fft{\del \cF(Q,P)}{\del P}=
\cF(Q,P)\,.\label{euler}
\eea

Moreover, the scalar charge $\Sigma$ can be expressed in terms of derivatives
of the mass with respect to $Q$ and $P$: \footnote{This was shown in
\cite{Cremonini:2021upd}, 
based on the result in \cite{gibkalkol} that the scalar
charge $\Sigma$, defined as in eqn (\ref{phiexp}), can also be written
as $\fft{\del M}{\del\phi_\infty}$, where generically the scalar 
$\phi$ is taken to be equal to $\phi_\infty$ at infinity.  As observed
in \cite{Cremonini:2021upd} (and adapted to our notation here), there is
a global shift symmetry of the EMD theory under which 
$\phi\longrightarrow \phi+
\phi_\infty$, $A_\mu\longrightarrow e^{-\ft12 a\phi_\infty}\, A_\mu$,
implying for the charges that $Q\longrightarrow e^{\ft12 a\phi_\infty}\, Q$ 
and $P\longrightarrow e^{-\ft12 a\phi_\infty}\, P$. Calculating
$\fft{\del}{\del\phi_\infty} M(e^{\ft12 a\phi_\infty}\, Q, e^{-\ft12 a\phi_\infty}\, P)$ in the shifted system, and then restoring $\phi_\infty=0$,
gives (\ref{Sig}).}
\bea
\Sigma = \ft12 a\, 
\Big( Q\fft{\del M}{\del Q} -P\, \fft{\del M}{\del P}\Big)\,.\label{Sig}
\eea
The long-range force $F$ between two identical extremal black holes is
given by 
\bea
r^2\, F= Q^2 +P^2 - \ft14 M^2 -\Sigma^2\,,
\eea
and, as we discussed in the previous section, this must 
vanish.  Plugging the expression (\ref{Sig}) for the scalar
charge into the 
equation $F=0$ gives a differential equation for $\cF(Q,P)$.  
Using the homogeneity relation (\ref{euler}) 
we can replace the $P\,\del\cF/\del P$ term in
the differential equation by $\cF - Q\, \del\cF/\del Q$, thus giving an 
ordinary differential equation in which $Q$ is viewed as 
the independent variable and
$P$ is just viewed as a fixed parameter.  If we furthermore define
$\tilde f(Q)$ by writing 
\bea
\cF(Q,P)= \sqrt{8 P Q}\, \tilde f(Q)\,,
\eea
we see that the equation following from requiring the vanishing of the force
becomes
\bea
Q^2+P^2 - 2 P Q\, \tilde f(Q)^2 - 8 a^2\, P\, Q^3\, (\del_Q\tilde f(Q))^2=0\,.
\eea

This equation can be simplified even more by defining a new independent variable $x$ in place of $Q$, where
\bea
Q= P\, e^{2ax}\,,\qquad -\infty \le x\le\infty\,.\label{Qx}
\eea
Finally, introducing $f(x)$, which is just $\tilde f(Q)$ with $Q$ replaced using
the definition (\ref{Qx}), we obtain the simple equation
\bea
\label{massDEQ}
\boxed{
f^2 + {f'}^2 = \cosh 2 a x\,,\label{feqn}
}
\eea
where $f'$ means $df(x)/dx$.  In view of the duality symmetry under the
the exchange of electric and magnetic charges, it is evident that the
required solution for $f(x)$ must obey
\bea
f(-x)=f(x)\,.\label{fsym}
\eea
When $x=0$, meaning $Q=P$, the dilaton becomes constant and the black
hole will just be the extremal Reissner-Nordstr\"om $Q=P$ dyon
for all values of $a$, with mass $M=\sqrt8\, Q$.  It follows therefore
that
\bea
f(0)=1\,,\qquad f'(0)=0\,.\label{f0fp0}
\eea

   As will be seen in appendix \ref{numericalsec}, our findings from studying
the solutions of eqn (\ref{feqn}) numerically are that, for all
values of $a>0$ the function $f(x)$ increases monotonically from
$f(0)=1$, as $x$ increases from 0.  The gradient $f'(x)$ also increases
monotonically from $f'(0)=0$ as $x$ increases from 0.  Of course because of
eqn (\ref{fsym}), $f(x)$ also increases monotonically as $x$ becomes 
increasingly negative, and $f'(x)$ becomes monotonically more negative as
$x$ becomes increasingly negative.

If we could solve equation (\ref{feqn}), it would give us $\cF(Q,P)$, and
hence we would have an explicit mass formula for extremal black holes for 
all values of $a$, namely
\bea
M=\cF(Q,P)= \sqrt{8QP}\, f(x) = \sqrt8\, P e^{ax}\, f(x)\,,\qquad \hbox{where}\quad
 x=\fft1{2a}\log\fft{Q}{P}\,.\label{massformula}
\eea
From eqn (\ref{Sig}), we can also express the scalar charge in terms of
$f(x)$, finding
\bea
\Sigma = \ft12\sqrt{8QP}\, f'(x)= \sqrt2\, P\, e^{ax}\, f'(x)\,.
\eea

  As a check, we can calculate what $f(x)$ is explicitly for the known
special cases.  Actually $a=0$ is a degenerate case in this parameterisation,
because of the redefinition (\ref{Qx}).  For the others, we see from
eqns (\ref{M3cases}) that we have
\bea
a=1:&& f(x) = \cosh x\,,\nn\\
a=\sqrt3: && f(x) = \Big( \cosh\fft{2x}{\sqrt3}\Big)^{3/2}
\,.
\eea
As can be verified, these expressions indeed satisfy eqn 
(\ref{feqn}).\footnote{In \cite{rasheed} Rasheed conjectured the mass 
formula $M=
2 (1+a^2)^{-1/2}\, (Q^b+P^b)^{1/b}$, with the constant $b$ given
by $b=(2\log 2)/(\log(2+2a^2))$, for
arbitrary values of $a$.
This would imply that our function $f(x)$ would become $f(x)=
(1+a^2)^{-1/2} \, 2^{(2-b)/(2b)}\, (\cosh a b x)^{1/b}$.  However, one can
easily verify that except for $a=1$ and $a=\sqrt3$, for which the Rasheed
conjecture does indeed (by construction) yield the known solutions,
for all other generic values of $a$ the $f(x)$ that results from the
Rasheed conjecture fails to satisfy eqn (\ref{feqn}).  This supports
a result in \cite{giant}, which also found that Rasheed's conjecture
could not be correct for general values of $a$.}

  Although the ODE in eqn (\ref{feqn}) looks very simple, it is not clear how
to solve it explicitly.
  We can, however, use the results obtained above in order to make some
observations about the force between two non-identical extremal 
black holes.  Suppose we have two such black holes, with charges 
$(Q_1,P_1)$ and $(Q_2,P_2)$.  The force $F_{12}$ between them is given  
by eqn (\ref{F120}),
and so, using the definitions given above,
\bea
r^2\, F_{12}= 2P_1 P_2 e^{a(x_1+x_2)}\, \Big[ \cosh a(x_1+x_2) - 
    f_1 f_2 - f_1' f_2'\Big]\,,\label{F12exp}
\eea
where $f_1$ means $f(x_1)$, $f_1'$ means $f'(x_1)$, and so on.  Thus once
$f(x)$ is known (for a given value of $a$), we can calculate the 
force between any pair of extremal black holes.

Note that equation (\ref{feqn}) provides a major computational 
simplification, compared to the
methods previously available to us for determining $M$.  Previously, we
would have to do a completely new numerical integration for each choice of
 $Q$ and  $P$.  Each such calculation required the
use of the shooting method to find the right choice for the expansion 
coefficient $e_2$ that gave a well-behaved solution with a regular black hole
horizon.   Now, by contrast, we simply have to carry out one numerical
calculation to obtain the result for $f(x)$.  With this result,
we can then immediately find the numerical result for $\cF(Q,P)$, and
hence for the mass, for any choice of $Q$ and $P$ that we like.  (Of course
in both approaches, we first pick a value for the dilaton coupling $a$.) 

\section{Perturbative Solutions Of $f(x)$}

 In the absence of an explicit closed-form solution for $f(x)$ for
general values of the dilaton coupling $a$, we can look at the solution
in various regimes where perturbative techniques can be applied.  
Before doing so, we should point out that the special choice of coupling 
 $a=1$ will play a crucial role in our discussion. 
 Indeed, when $a=1$  the extremal solutions are BPS, and the long distance force between any two of them is always zero.
 As we will see further below, the long distance nature of the force (i.e., whether it is repulsive or attractive) will be correlated with whether 
 $a$ is above or below this special value.

\subsection{Perturbations around $a=1$}\label{anear1sec}

  We know that when $a=1$, the solution to the equation (\ref{feqn}) is
$f(x)=\cosh x$.  We may now consider perturbing around $a=1$, by writing
\bea
a=1+\ep\,,\qquad f(x)= \cosh x + \ep \,u_1(x) + \ep^2\, u_2(x) +\cdots\,.
\eea
At order $\ep$, we therefore find that $u_1(x)$ must satisfy
\bea
2\sinh x\, u_1' + 2 \cosh x\, u_1 - 2 x \sinh 2 x=0\,.
\eea
The solution that is regular at $x=0$ is given by
\bea
u_1= \fft{2 x \cosh 2 x - \sinh 2x}{4 \sinh x}\,.\label{u1sol}
\eea
Thus for $a=1+\ep$ we have the solution
\bea
f(x) = \cosh x + \ep\, \fft{2 x \cosh 2 x - \sinh 2x}{4 \sinh x} +
   {\cal O}(\ep^2)\,.\label{fforanear1}
\eea

One could also consider an expansion around the other exactly-solvable
case, $a=\sqrt3$, but this would be of less interest than the expansion
around $a=1$.  
Indeed, as we mentioned above, the case $a=1$ has special significance 
because it supports
BPS extremal solutions, whose long range interactions vanish independently of the values of the charges of each black hole solution.
Thus, by probing the regime close to $a=1$ we shall be able to see the
transition from having attractive forces when $a<1$ and repulsive forces
when $a>1$.

\subsection{Perturbative expansion for small $x$ ($Q$ close to $P$)}\label{smallxsec}

First, we consider a 
perturbative expansion for $f(x)$ in the regime where $x$ is
small.  In view of the relation $Q=P\, e^{2ax}$ in eqn (\ref{Qx}), this
corresponds to the situation where $Q$ is close to $P$.  

   We can look for the perturbative solution for $f(x)$ in eqn (\ref{feqn}),
by expanding $f(x)$ in the form
\bea
f(x)= \sum_{n\ge 0} a_{2n}\, x^{2n}\,.\label{fexp}
\eea
(The required solutions should be symmetric under $x\longrightarrow - x$,
since the black hole metrics must be invariant under exchanging the electric
and magnetic charges.)  Writing $a$ in terms of a parameter $k$, where
\bea
a^2=\ft12 k(k+1)\,,\label{akrel}
\eea
and expanding eqn (\ref{feqn}) in powers of $x$, one
can solve for the coefficients in eqn (\ref{fexp}), finding
\bea
a_0 &=& 1\,,\qquad a_2= \fft{k}{2}\,,\qquad 
a_4= \fft{k^2(2k^2 + 4k -1)}{24 (4k+1)}\,,\nn\\
a_6 &=& \fft{k^3 (24 k^5 + 64 k^4 + 52 k^3 - 18 k^2 + 34k +19)}{
            720 (4k+1)^2 (6k+1)}\label{smallxcoeffs}
\eea
and so on.  In fact, this small-$x$ expansion can be seen to be in
agreement 
with an expansion for the mass of extremal EMD black holes that was
obtained in eqn (4.9) of \cite{giant} for the case where $Q$ and $P$
are nearly equal (the $\epsilon$ parameter in \cite{giant} is equivalent to
$2x$ in our expansion here).  
We have verified the agreement up to order $x^{10}$ 
in the expansion (\ref{fexp}).

    For our purposes later in the paper, it will suffice to keep the terms
just up to order $x^2$ in the expansion, and so we have the small-$x$
expansion
\bea
f(x) = 1 + \ft12 k\, x^2 + {\cal O}(x^4)\,.\label{fsmallx}
\eea

\subsection{Large-$x$ expansion (large hierarchy between charges)}

  When $x$ is very large and positive, so that there is a large hierarchy between $Q$ and $P$, eqn (\ref{feqn}) becomes approximately
\bea
f^2 + {f'}^2 =\ft12 e^{2a x}\,.
\eea
Thus $f(x)$ for large positive $x$ takes the form
\bea
f(x) \approx \fft{e^{a x}}{\sqrt{2(1+a^2)}}\,.\label{xlargep}
\eea
Since we must have $f(-x)=f(x)$, the solution when $x$ is large and
negative takes the form
\bea
f(x) \approx \fft{e^{-a x}}{\sqrt{2(1+a^2)}}\,.\label{xlargen}
\eea

\section{Force Between Non-identical Extremal Black Holes}

We are now ready to examine the main issue we want to address in this paper, which is 
whether long range forces between non-identical black holes display any generic features that are tied to the structure of the theory and of the scalar couplings. 
As we are about to see, the behavior of long range interactions in the EMD model is 
controlled in a simple way by the dilatonic coupling $a$. In particular,  the sign of the force 
between distinct extremal black holes -- whether it is attractive or repulsive -- is correlated with whether $a>1$ or $a<1$.

  In eqn (\ref{F12exp}) we gave a general expression for the long-range
force between two non-identical extremal static 
black holes in the EMD theory, for
an arbitrary value of the dilaton coupling $a$, which we
repeat here for convenience:
\bea
r^2\, F_{12}= 2P_1 P_2 e^{a(x_1+x_2)}\, \Big[ \cosh a(x_1+x_2) -
    f_1 f_2 - f_1' f_2'\Big]\,.\label{F12exp2}
\eea
The expression is
written in terms of the electric and magnetic charges $(Q_1,P_1)$ and
$(Q_2,P_2)$ for the two black holes, with $Q_1$ and $Q_2$ written as
\bea
\label{QPx}
Q_1= e^{2a x_1}\, P_1\,,\qquad Q_2=e^{2a x_2}\, P_2\,.
\eea

  Note that the force $F(Q_1,P_1;Q_2,P_2)$ between extremal black holes
with charges $(Q_1,P_1)$ and $(Q_2,P_2)$ necessarily has the homogeneity
property that\footnote{This homogeneity accounts, in particular,  
for the fact, mentioned in footnote \ref{identfoot}, 
that the zero-force property for two extremal black holes
that have identical sets of charges also holds if the set of charges of one 
of the black holes is an overall multiple of the set of charges of the 
other black hole.}
\bea
F(\lambda_1\, Q_1,\lambda_1\, P_1;\lambda_2\, Q_2,\lambda_2\, P_2)
=\lambda_1\, \lambda_2\, F(Q_1,P_1;Q_2,P_2)\,.\label{scalings}
\eea
Thus when looking at the force as a function of the
charges  $(Q_1,P_1,Q_2,P_2)$, there is really only a two-dimensional 
parameter space of of non-trivially 
inequivalent configurations to explore, rather than
the three-dimensional parameter space one might naively have expected.
That is to say, there is not only the obvious overall scaling, 
under which $F(Q_1,P_1;Q_2,P_2)$ would scale by $k^2$ if all
four charges were scaled by $k$, but there are the two separate, independent
scalings of the two sets of charges $(Q_1,P_1)$ and $(Q_2,P_2)$, 
as seen in eqn (\ref{scalings}).  

   This double scaling homogeneity is seen in the expression (\ref{F12exp2})
for the force between the two black holes, with the product of the two
magnetic charges appearing in the prefactor.  The non-trivial charge-dependence
of the force (i.e. dependence that is not merely taking the form of an
overall scaling of the force) is then encapsulated by the two 
parameters $x_1$ and $x_2$, which characterise the ratio of the
electric to magnetic charge for each of the black holes through the
relations (\ref{QPx}).

Since the function
$f(x)$ is not known explicitly for general values of $a$, we would have
to determine it numerically, by solving eqn (\ref{feqn}), in order to
explore the full $(x_1,x_2)$ space of non-trivial parameter values. 
We can however
obtain some analytical results in certain regimes, as we shall now discuss.

\subsection{Force between extremal black holes with $a$ 
 near to 1}\label{anear1force}

  We saw in section \ref{anear1sec} that if we consider values of
$a$ close to $a=1$, by writing $a=1+\ep$, 
 the solution to eqn (\ref{feqn}) up to order $\ep$ is given by eqn (\ref{fforanear1}).
Substituting this expression for $f(x)$ into eqn (\ref{F12exp2}) gives
the force
\bea
r^2\, F_{12} = \ep\, P_1\, P_2\, e^{a(x_1+x_2)}\, \sinh(x_1-x_2)\, 
 \Big[ H(x_1)- H(x_2)\Big] + {\cal O}(\ep^2)\,,\label{F12ep}
\eea
where we have defined 
\bea
H(x)= \fft{\sinh 2x - 2 x}{2\sinh^2 x}\,.
\eea
The function $H(x)$ is monotonically increasing from $H(-\infty)=-1$
to $H(\infty)=+1$, with $H(-x)=-H(x)$.  Therefore we see that the coefficient
of $\ep$ in eqn (\ref{F12ep}) is always non-negative, for all $x_1$ and
$x_2$.  It follows that $F_{12}$ is positive (repulsive) when $\ep$ is small
and positive (i.e. $a>1$) and the force is negative (attractive) when $\ep$
is small and negative (i.e. $a<1$).

   This calculation for the case where $a$ is close to 1 suggests that 
it may more generally be true that the force between non-identical 
extremal static EMD black holes is always positive (repulsive) when $a>1$,
and always negative (attractive) when $a<1$.  We shall now establish 
further evidence to support this proposition, by considering other 
regimes where we can perform analytical calculations.

\subsection{Force between extremal black holes with $x$ small}

   Since $Q=e^{2ax}\, P$, the black holes with $Q$ close to $P$
correspond to the case where $x$ is small.  Using the small-$x$ expansion for
$f(x)$ that we obtained in section \ref{smallxsec}, we may substitute 
$f(x)$ given by eqns (\ref{fexp}) and (\ref{smallxcoeffs}) into the
expression (\ref{F12exp2}) for the force between two such black holes, finding
\bea
r^2\, F_{12} = \ft12 P_1 P_2\,\big[ k(k-1)\, (x_1-x_2)^2 +
\cdots\big]\,,
\eea
where the ellipses denote terms of higher
than quadratic order in the small quantities $x_1$ and $x_2$.  Since
$a^2=\ft12 k(k+1)$, we see that in this regime 
(where the $x$ parameters are small) 
the force between two extremal black holes is again 
positive whenever
$a>1$ and negative whenever $a<1$. 

\subsection{Force between extremal black holes with $x$ large}

Recall that large $x$ corresponds to a large hierarchy between the associated charges.
 There are several cases of interest that we may consider here:
\begin{itemize}

\item 
\noindent{\underline{\bf {$x_1$ large and positive, $x_2=0$}}:}

 Using the expression (\ref{xlargep}) for $f(x_1)$, and eqns 
(\ref{f0fp0}) for $f(x_2)$, we find that the force (\ref{F12exp2})
between the two black holes becomes
\bea
r^2\, F_{12} \sim P_1\, P_1\, e^{2a x_1}\, 
            \Big[1-\sqrt{\fft{2}{1+a^2}}\,\Big]
\eea
as $x_1$ becomes large.  As in the previous specialisations we considered,
here too the force will be positive if $a>1$ and negative if $a<1$.

\item
\noindent{\underline{\bf {$x_1$ large and positive, $x_2$ large 
and negative}}:}

Let us take
\bea
x_1= k+\lambda\,,\qquad x_2=-k\,,
\eea
where $k$ is large and positive, and $\lambda$ is held fixed.  Using
eqn (\ref{xlargep}) for $f(x_1)$ and eqn (\ref{xlargen}) for $f(x_2)$,
we find that the force in eqn (\ref{F12exp2}) becomes
\bea
r^2\, F_{12} \sim P_1\, P_2\, e^{2a(k+\lambda)}\, \fft{a^2-1}{a^2+1}\,.
\eea
In this regime also, the force is positive when $a>1$ and 
negative when $a<1$.

\item

\noindent{\underline{\bf {$x_1$ and $x_2$ large and positive}}:}
\medskip

Taking $x_1=k+\lambda$ and $x_2=k$, and sending $k$ to infinity while holding
$\lambda$ fixed, just gives the result that $F_{12}$ goes to zero, if we 
work at the level of approximation of the expression (\ref{xlargep}) for
$f(x)$ at large $x$.  This is not too surprising, since when $x_1$ and
$x_2$ are both becoming very large, it means that the electric charges
of the two black holes are completely overwhelming their magnetic
charges.  Thus with their charges becoming effectively (after rescaling)
of the form $(\tilde Q_1,0)$ and $(\tilde Q_2,0)$, it means that the charges
of one black hole are just a multiple of the charges of the other, and so,
just as for identical black holes, the force will be zero.

\end{itemize}

\subsection{Force between nearly identical extremal black holes}

Next, we take two extremal black holes with charges $(Q_1,P_1)$ and 
$(Q_2,P_2)$, for the case
\bea
Q_1= P_1\, e^{2ax}\,,\qquad Q_2= P_2\, e^{2 a (x+\ep)}\,,
\eea
with $\ep$ a perturbatively small parameter.
Substituting into the expression (\ref{F12exp2}) for the force 
between the black holes and expanding to order $\ep^2$ we find, after making
use of eqn (\ref{feqn}) and its first two derivatives, that
\bea
r^2\, F_{12} = \ep^2\, e^{2 a x}\, P_1\, P_2\, \Big( a^2\cosh 2 a x-
  f(x)\, f''(x) - f'(x)\, f'''(x) \Big) + {\cal O}(\ep^3)\,.
\label{F12inf}
\eea
 Thus, the sign of the force is governed by the factor
\bea
G=a^2\cosh 2 a x- f(x)\, f''(x) - f'(x)\, f'''(x)\label{FGdef} \, .
\eea
By making use of the equation (\ref{feqn}) for
$f(x)$, and its derivatives, one can recast the expression for $G$ into the 
form
\bea
G= -a^2\cosh 2 a x + {f'(x)}^2 + {f''(x)}^2\,.\label{Gexp2}
\eea
One can also rewrite the expression for $G$ in such as way that all
derivatives of $f$ are eliminated, by making use of (\ref{feqn}) and
its derivatives.  This gives
\bea
G= (1-a^2)\,\cosh 2ax +\fft{a^2\,\sinh^2 2ax}{\cosh 2ax - f(x)^2} -
\fft{2 a f\, \sinh 2 a x}{\sqrt{\cosh 2ax - f(x)^2}}\,.\label{Gexp3}
\eea

 If one solves eqn (\ref{feqn}) numerically for some specified value of $a$,
and then substitutes into (\ref{F12inf}) or one of the
expressions for $G$ above, one finds that the force is
positive when $a>1$ and negative when $a<1$, in accordance with previous
expectations.  From the standpoint of numerical accuracy, it is
advantageous to use the expression for $G$ in eqn (\ref{Gexp3}), where
derivatives of the numerically-determined function $f(x)$ are not needed.

\subsection{Force inequalities}

We conclude this section by mentioning that the force is constrained to obey certain bounds.
Consider the identities
\bea
(f_1 \pm f_2)^2 + (f_1'\pm f_2')^2 &=& f_1^2 + {f_1'}^2 +
   f_2^2 + {f_2'}^2  \pm 2 (f_1 f_2 + f_1' f_2')\nn\\
&=& \cosh 2 a x_1 + \cosh 2 a x_2 \pm 2 (f_1 f_2 + f_1' f_2')\,,
\eea
where we have used eqn (\ref{feqn}) in order to obtain the second line.  Thus
we see that
\bea 
-\cosh 2a x_1 - \cosh 2 a x_2 \le 2 (f_1 f_2 + f_1' f_2') \le
\cosh 2 a x_1 + \cosh 2 a x_2\,.
\eea
In consequence, it follows from eqn (\ref{F12exp}) that the force
$F_{12}$ between two extremal black holes obeys the bounds
\bea
-\sinh^2\fft{a(x_1-x_2)}{2}
  \le  \hat F_{12}\le \cosh^2\fft{a(x_1-x_2)}{2}\,,
\eea
where we have defined $\hat F_{12}$ to be the positive multiple of the
force $F_{12}$ given by
\bea
r^2\, F_{12}= 4 P_1 P_2 e^{a(x_1+x_2)}\, \cosh a(x_1+x_2)\, \hat F_{12}\,.
\eea
The inequalities are not powerful enough to provide useful information
about the sign of the force, but they do place constraints on the
magnitude of the force as a function of $a$ and the charges.

\section{Binding Energy}

Now that we have seen evidence of a direct connection between the range of the coupling $a$ and the long distance nature 
of the force, we want to examine what happens to the binding energies in the theory. The naive expectation is that they should exhibit 
the same kind of behaviour, and indeed they do. 
As we will show below, depending on whether $a>1$ or $a<1$  the binding energies will be positive or negative.
Moreover, their sign will be correlated with the convexity or concavity of the surface $M=\cF(Q,P)$
which describes the dependence of the mass on the charges.

  Consider two extremal EMD black holes, with charges $(Q_1,P_1)$ and
$(Q_2,P_2)$.  We may also consider a ``composite'' extremal black hole,
with charges $(Q_1+Q_2,P_1+P_2)$, and then define a notion of binding
energy $\Delta M$ as the difference between the mass $\hat M$ of the composite
black hole and the sum of the masses $M_1$ and $M_2$ 
of the two constituents\footnote{We are assuming for simplicity that the composite state is extremal. However, it doesn't have to be.}:
\bea
\Delta M = \hat M- M_1 -M_2\,.
\label{DeltaM}
\eea
Since we have introduced the function $\cF(Q,P)$ as giving the mass of an 
extremal black hole with charges $(Q,P)$ (see eqn (\ref{MfromcF})), we 
therefore have
\bea
\Delta M =\cF(Q_1+Q_2,P_1+P_2) -\cF(Q_1,P_1) - \cF(Q_2,P_2)\,.\label{DMcF}
\eea

   It would seem intuitively reasonable to expect that if the binding
energy $\Delta M$ is positive, then it would be energetically favourable
for the composite extremal black hole to separate into its
two component black holes.  In other words, we might expect that if
$\Delta M$ is positive, then there should be a repulsive force between
the two constituent black holes with charges $(Q_1,P_1)$ and
$(Q_2,P_2)$.  On the other hand, if $\Delta M$ is negative, we might
expect that there would be an attractive force between the two constituent
black holes.

To examine the structure of $\Delta M$  it turns out to be convenient 
 to define the two-dimensional energy 
surface in $\R^3$ with $(x,y,z)$
coordinates $(Q,P,\cF(Q,P))$.  The sign of $\Delta M$
defined in eqn (\ref{DMcF}) is then related 
to the convexity or concavity of the surface.
First, note that the homogeneity property $\cF(\lambda Q, \lambda P)=
\lambda\, \cF(Q,P)$ means that we can rewrite eqn (\ref{DMcF}) as
\bea
\Delta M =2 \cF\big(\fft{Q_1+Q_2}{2},\fft{P_1+P_2}{2}\big) 
-\cF(Q_1,P_1) - \cF(Q_2,P_2)\,.\label{DMcF2}
\eea
The two extremal black holes with charges $(Q_1,P_1)$ and
$(Q_2,P_2)$ define two points on the energy surface, namely
\bea
(Q,P,\cF(Q,P))\quad \hbox{and} \quad  (Q',P',\cF(Q',P')) \,.
\eea
If we draw a straight line in $\R^3$ joining these two points, its
midpoint will lie at
\bea
\Big(\fft{Q+Q'}{2},\fft{P+P'}{2}, \fft{\cF(Q,P)+\cF(Q',P')}{2}
\Big)\label{midpoint}
\eea
in $\R^3$, and  in general it will \emph{not} lie on the energy surface.
Now consider the point in $\R^3$, which {\it does} lie on the two-dimensional
surface, whose coordinates are
\bea
\Big(\fft{Q+Q'}{2},\fft{P+P'}{2}, \cF(\fft{Q+Q'}{2},\fft{P+P'}{2})\Big)\,.
\label{midonsurface}
\eea
Since the $x$ and $y$ coordinates of the two points (\ref{midpoint}) and
(\ref{midonsurface}) are the same, the two points sit vertically one above
the other.  

  We can now see that if the energy surface defined by the equation 
$M=\cF(Q,P)$
is {\it convex}, then the point (\ref{midonsurface}) will lie 
{\it above} the point (\ref{midpoint}).  On the other hand,
if the energy surface 
is {\it concave}, then the point (\ref{midonsurface}) will lie
{\it below} the point (\ref{midpoint}).  In other words, we have
\bea
\hbox{Convex}:&&\qquad \cF(\fft{Q+Q'}{2},\fft{P+P'}{2}) >
\fft{\cF(Q,P)+\cF(Q',P,)}{2}\,,\nn\\
\hbox{Concave}:&&\qquad \cF(\fft{Q+Q'}{2},\fft{P+P'}{2}) <
\fft{\cF(Q,P)+\cF(Q',P,)}{2}\,.
\eea
Thus, from eqn (\ref{DMcF2}), it follows that we have:
\bea
\hbox{Convex}:&&\qquad \Delta M>0\,,\nn\\
\hbox{Concave}:&&\qquad \Delta M <0\,,\label{DMsign}
\eea
tying the sign of the binding energy to the shape of the energy surface.

\subsection{Binding energy for nearly identical extremal black holes}

   One can also look at the relation between the sign of the binding energy
and the concavity or convexity of the energy surface at the infinitesimal
level.  Consider two extremal black holes with charges $(Q,P)$ and
$\big((1+\alpha)\, Q, (1+\beta)\, P)$, where $\alpha$ and $\beta$ are 
infinitesimal.  Calculating the binding energy $\Delta M$ given in 
 eqn (\ref{DMcF2}), we find, up to quadratic order in $\alpha$ and $\beta$,
\bea
\Delta M= -\ft14(\alpha^2 Q^2\, \del_Q^2  + \beta^2 P^2 + 2\alpha\beta Q P)\,
\cF(Q,P)\,.\label{DMinf}
\eea
Thus again we see that if the energy surface is locally 
convex then $\Delta M$ is 
positive, while if the energy surface is locally concave then $\Delta M$ is 
negative.

  If we use eqn (\ref{massformula}) to write $\cF(Q,P)$ in terms of $f(x)$,
as 
$\cF(Q,P)=\sqrt8\, P\, e^{ax}\, f(x)$ (recalling that
$Q=e^{2 ax}\, P$), then the expression (\ref{DMinf}) for the binding
energy becomes
\bea
\Delta M= \fft{P\, e^{ax}}{4\sqrt2\, a^2}\, (\alpha-\beta)^2 \,
  [a^2 f(x) - f''(x)]\,.\label{DMinf2}
\eea
Thus, the sign of the binding energy is governed by the sign of $a^2 f(x)-f''(x)$.

   Using our expansion in eqn (\ref{fforanear1}) for the case when
$a=1+\ep$, we see that up to linear order in $\ep$ 
\bea
a^2 f(x)-f''(x) = \fft{\ep}{2\sinh^3x}\, (\sinh2x -2x)\,.
\eea
Since $(\sinh 2x - 2x)$ is positive (negative) when $x$ is positive (negative), it follows that $a^2 f(x)-f''(x)$ is
positive when $a=1+\ep$ with $\ep$ positive, and negative when $\ep$ is
negative.  In other words, at least in the regime where $a$ is close to
1, the sign of the binding energy is indeed correlated with the sign of
the force between extremal black holes that we saw previously.

  As another check, we can look at the leading-order term in $a^2 f(x)-f''(x)$
when $x$ is small, using the expansion for $f(x)$ in eqns (\ref{akrel}) and 
(\ref{smallxcoeffs}).  We find
\bea
a^2 f(x) - f''(x) = \ft12 k(k-1) + {\cal O}(x^2)\,,
\eea
which, since $a^2=\ft12 k(k+1)$, shows that in this small-$x$ regime 
$\Delta M$ is again positive when $a>1$ and negative when $a<1$.

To probe the entire parameter space we have to resort to
numerical methods in order to solve for $f(x)$ for some chosen value of
$a$.  Our numerical results indicate that our observations above are robust, i.e. they show that $a^2 f(x) - f''(x)$, and
hence the binding energy, is always positive when $a>1$ and negative when $a<1$.

\subsection{Binding energy between extremal black holes with $a$ near to 1}
\label{DMseca1}

 The general expression in eqn (\ref{DMcF}) for the binding energy
between extremal black holes with charges $(Q_1,P_1)$ and $(Q_2,P_2)$
becomes, upon using the expression (\ref{massformula}) for 
$\cF(Q,P)$ in terms of $f(x)$,
\bea
\fft{\Delta M}{\sqrt 8}= (P_1+P_2) e^{a\hat x}\, f(\hat x) -
P_1\, e^{a x_1}\, f(x_1) - P_2\, e^{a x_2}\, f(x_2)\,,\label{DMgen}
\eea
where 
\bea
Q_1 = e^{2 a x_1}\, P_1\,,\qquad Q_2= e^{2 a x_2}\, P_2\,,\qquad
(Q_1+ Q_2) = e^{2 a \hat x}\, (P_1+P_2)\,.\label{QPxdefs}
\eea
The parameter $\hat x$ for the composite extremal black holes
is thus determined in terms of $P_1$, $P_2$, $x_1$ and $x_2$ by the equation
\bea
(e^{2 a \hat x} - e^{2a x_1}) \, P_1 + (e^{2 a  \hat x} - e^{2a x_2})\, 
P_2=0\,.\label{xhateqn}
\eea
We may assume without loss of generality that $x_2 < x_1$.  It then 
follows from
eqn (\ref{xhateqn}) that $\hat x$ must satisfy
\bea
x_2 < \hat x < x_1\,.\label{xbounds}
\eea
Rather than viewing $\hat x$ as a derived quantity after $P_1$, $P_2$,
$x_1$ and $x_2$ are specified, we may instead
view $P_1$, $x_1$, $\hat x$ and $x_2$, subject to eqn (\ref{xbounds}), as
the four parameters characterising the parameter space of
the two extremal black holes.  Since $P_1$ will then be merely an overall
multiplicative factor in the expression for the binding energy, we have
just three non-trivial parameters to consider, 
namely $x_1$, $\hat x$ and $x_2$, subject
again to (\ref{xbounds}).

   Now, let us consider the situation where $a$ is close to 1, so as 
before we write $a=1+\ep$.  To linear order in $\ep$, the function
$f(x)$ is given by eqn (\ref{fforanear1}).  We then find that to linear
order in $\ep$, the binding energy given in eqn (\ref{DMgen}) becomes
\bea
\Delta M= \fft{\ep \,P_1\, e^{a x_1} \,W}{\sqrt8 \, \sinh x_1\, \sinh x_2\, 
\sinh \hat x\, \sinh(\hat x -x_2)}\,,\label{DMneara=1}
\eea
where
\bea
W(x_1,\hat x,x_2) &=& 
(\hat x-x_2)\, \sinh 2 x_1 + (x_1-\hat x)\, \sinh 2 x_2 +
  x_2\, \sinh 2(x_1 - \hat x) \nn\\
&& + x_1\, \sinh2(\hat x-x_2) -
  \hat x\, \sinh2(x_1-x_2) -(x_1-x_2)\, \sinh 2\hat x\,.\label{Wdef}
\eea
It can easily be verified numerically, 
by testing numerous random choices for triples $(x_1, \hat x, x_2)$
obeying the inequality (\ref{xbounds}), that the coefficient of $\ep$ in
eqn is indeed
always positive.  This indicates that the binding energy is positive when
$\ep>0$ and negative when $\ep<0$.  

   We have also constructed an analytical proof of the positivity of the
coefficient of $\ep$ in eqn (\ref{DMneara=1}).  Since the proof is a little 
intricate, we have relegated it to appendix \ref{DMappendix}.

\section{A Two-Dimensional Picture}

Next, we would like to examine in some more detail the connection between  
the shape of the energy surface and the sign of the binding energy, which we mentioned in (\ref{DMsign}).
   Owing to the homogeneity $\cF(\lambda Q,\lambda P)=\lambda\, \cF(Q,P)$
of the mass function, another way to picture the properties
of the energy surface is divide out the equation $M=\cF(Q,P)$ by $M$ and
hence obtain
\bea
\cF(u,v)=1\,,\qquad \hbox{where}\quad u=\fft{Q}{M}\,,\qquad 
 v= \fft{P}{M}\,.
\eea
The shape of the curve $\cF(u,v)=1$ in the $(u,v)$ plane captures 
the characteristics of the energy surface.  From the expressions
$M=\sqrt{8QP}\, f(x)$ and $Q=P e^{2ax}$ in section \ref{MassDEQ} we see
that the curve $\cF(u,v)=1$ can be written parametrically as
\bea
u(x) = \fft{e^{ax}}{\sqrt8\, f(x)}\,,\qquad 
   v(x) = \fft{e^{-ax}}{\sqrt8\, f(x)}\,. \label{uxvx}
\eea
In the positive quadrant that we are considering, the $\cF(u,v)=1$
curves run from a point on the positive $v$ axis (corresponding to $x=-\infty$)
to a point on the $u$ axis (corresponding to $x=+\infty$). 

  As can be seen in the examples plotted in appendix \ref{numericalsec}, the
$\cF(u,v)=1$ curves are concave if $a>1$ and convex if $a<1$.   Note
that when the curve $\cF(u,v)=1$ is concave, the energy surface
$M=\cF(Q,P)$ is convex, and vice versa.  This is just an inherent feature
of the two different but equivalent 
ways of characterising the same information contained
in the mass function $\cF(Q,P)$. This may be seen explicitly as follows:

   Consider first the case where the energy surface is {\it convex}, which
means that $\Delta M>0$ (see eqn (\ref{DMsign})).  By eqn (\ref{DMcF}),
this means that 
\bea
\Delta M= \cF(Q_1+Q_2,P_1+P_2) - \cF(Q_1,P_1) -\cF(Q_2,P_2) >0\,.
\eea
Since $\cF(Q_1,P_1)=M_1$ and $\cF(Q_2,P_2)=M_2$, it follows that
\bea
\Delta M= 
(M_1+M_2) \Big[ \cF\Big(\fft{Q_1+Q_2}{M_1+M_2}, \fft{P_1+P_2}{M_1+M_2}\Big)
  -1\Big]>0\,,\label{cFineq}
\eea
where we have used the homogeneity property $\cF(\lambda Q,\lambda P)=\lambda 
\,\cF(Q,P)$.  The points
$(u_1,v_1)$ and $(u_2,v_2)$ in the $(u,v)$ plane lie on the
curve $\cF(u,v)=1$, where
\bea
u_1=\fft{Q_1}{M_1}\,, \qquad v_1=\fft{P_1}{M_1}\,,\qquad
u_2=\fft{Q_2}{M_2}\,, \qquad v_2=\fft{P_2}{M_2}\,.
\eea
Now define the point $(\bar u,\bar v)$, where
\bea
\bar u= \fft{Q_1+ Q_2}{M_1+M_2}\,,\qquad \bar v= \fft{P_1+P_2}{M_1+M_2}\,.
\label{ubarvbar}
\eea
This point 
lies on the straight line joining $(u_1,v_1)$ to $(u_2,v_2)$, since it is
of the form
\bea
(\bar u,\bar v)= (u_1,v_1) +\lambda (u_2-u_1,v_2-v_1)
\eea
with 
\bea
\lambda = \fft{M_2}{M_1+M_2}\,.\label{lamdef}
\eea
Since eqn (\ref{lamdef}) implies that 
$\lambda$ lies somewhere between 0 and 1, it follows that the
point $(\bar u, \bar v)$ must lie {\it between} $(u_1,v_1)$ and $(u_2,v_2)$.
It follows from eqn (\ref{cFineq}) and the definitions (\ref{ubarvbar}) 
that when $\Delta M>0$ we must
have $\cF(\bar u,\bar v)>1$.  In other words, the point $(\bar u,\bar v)$
in the $(u,v)$ plane lies {\it outside} the curve 
$\cF(u,v)=1$.\footnote{Note that because of the homogeneity of the 
function $\cF(u,v)$, the curve $\cF(u,v)=k$ is a scaled version of the
curve $\cF(u,v)=1$, with the points $(u,v)$ on the curve scaled to
$(ku,kv)$.}   That is
to say, the curve $\cF(u,v)=1$ must be {\it concave}.

  In summary, we have shown that if the energy surface is {\it convex}
then the curve $\cF(u,v)=1$ is {\it concave}.  Of course the converse
is true also.  

Finally,  using the standard formula for the radius of curvature of a
parametric curve $\big(u(x),v(x)\big)$, namely
\bea
K= \fft{u'\, v'' - v'\, u''}{({u'}^2 + {v'}^2)^{3/2}}\,,
\eea
we see from (\ref{uxvx}) that here
\bea
u'\, v''- v'\, u'' = \fft{2 a}{f^3(x)}\, \Big(a^2\, f(x) -
f''(x)\Big)\,.\label{curveval}
\eea
The radius of curvature $K$ is therefore a positive quantity multiplied by
eqn (\ref{curveval}).  Thus the sign of $a^2\, f - f''$ provides
the criterion for determining the
convexity or concavity at each point along the curve.  As is to
be expected from our previous discussions of convexity and the sign 
of the binding energy, the quantity $a^2\, f - f''$ is the same one 
that governed the sign of the binding energy in eqn (\ref{DMinf2}).
Thus, we see a direct correlation between the range of the coupling $a$ (larger or smaller than one) and the curvature of the energy surface.

\section{Conclusions}

One way in which gravity being weak affects low energy EFTs is through the behavior 
of long range interactions in the theory. 
The latter can also be mediated by moduli -- massless scalars fields with vanishing potentials -- which are 
ubiquitous in string theories.
Thus, despite their simplicity, long range forces can potentially carry useful information about 
quantum gravity signatures on low energy physics. 
This observation was one of the motivations behind the RFC.
However, thus far essentially all of the work on probing long range interactions in the context of the Swampland program has been restricted to self-forces. It is natural to wonder if interactions between non-identical states can teach us new lessons, beyond what can be accessed by inspecting self-forces.

In this paper, motivated by the results of \cite{Cremonini:2022sxf}, we have examined 
long range forces and binding energies between non-identical static 
extremal dyonic black hole solutions to the simple EMD model (\ref{emdlag}).
These solutions are known explicitly only for three special
values of the dilaton coupling constant, namely $a=0$, $a=1$ and $a=\sqrt3$.
For generic values of $a$ the black hole solutions can only be obtained
numerically.  Although in principle the extremal black hole mass and 
the scalar charge must be determined purely in terms of the electric and
magnetic charges, it would be very laborious to explore the parameter
space of the solutions, in order then to calculate the force between
two black holes, by such numerical methods.  

    A key result in this paper is 
that we were able to find the simple first-order ordinary differential 
equation (\ref{feqn}) for a function $f(x)$, from which the mass and 
scalar charge can then be calculated.  Although we have not been able to 
find the explicit solution to this equation (except at $a=1$ and $a=\sqrt3$),
it is very easy to solve it numerically, performing just one
numerical integration for any chosen value of $a$.  Having obtained the
numerical solution for the chosen value of $a$, all information about the
mass and the scalar charge is then accessible.   In various special
cases, such as when the ratio of $Q$ to $P$ is very large or very small, 
or when $Q$ is very close to $P$, or when the dilaton coupling $a$ is
very near to 1, one can solve for $f(x)$ by perturbation methods.

   We were then able to identify a number of novel features in the
behaviour of the mass function, and the 
force between non-identical extremal black holes.
First of all,  the range of $a$ determines the shape of the 
surface relating the extremal mass of each black hole solution to its electric and magnetic charges, $M=\cF(Q,P)$.
In particular, 
\begin{eqnarray}
&& a>1 \quad \Rightarrow \quad \text{M is convex} \\
&& a<1 \quad \Rightarrow \quad \text{M is concave} \, .
\end{eqnarray}
Moreover,  the sign of the long range force between distinct extremal black holes is 
also correlated with the range of the coupling, meaning that 
\begin{eqnarray}
&& a>1 \quad \Rightarrow \quad \text{repulsive force} \\
&& a<1 \quad \Rightarrow \quad \text{attractive force} \, ,
\end{eqnarray}
with the borderline case $a=1$, for which the corresponding dyonic 
black holes are BPS, always giving a vanishing force.
Finally, as naively expected, the sign of the binding energy 
$\Delta M$ is correlated with the behavior of the long-range force.

A caveat of our discussion about the binding energy 
is that our analysis assumes  -- for simplicity -- that both the 
initial and final states are extremal black holes. This, of course, 
doesn't have to be the case.
Indeed,  the naive correlation between the long distance force and the ability to form bound states might cease to exist if the final states are non-extremal. It would be interesting to better understand this case, and what the implications of an attractive or repulsive force would be in that case.

A natural question is whether our results can be connected with the RFC -- especially its strongest formulation, 
in terms of strongly self-repulsive states.
It is interesting that 
our results indicate that in the EMD theory with $a<1$, all extremal dyonic black holes attract all other such solutions (assuming they carry different charges).  
This statement in itself is compatible with the RFC which, after all, doesn't require strongly self-repulsive multi-particle states to be black holes. 
However, it does raise the question of whether there is a more fundamental 
distinction between theories with $a>1$ or $a<1$, and if so, what is its origin. 
Independently of the RFC, it would be valuable to identify sharp criteria for the existence or absence of bound states.
A more detailed understanding of binding energies and bound states is not only relevant to flat space but also to Anti-de Sitter space, where the concept of repulsive forces needs to be  
expressed in an entirely different way (see e.g. \cite{Aharony:2021mpc,Andriolo:2022hax,Orlando:2023ljh}).
We wonder if some of the features we have identified have an analog description in Anti-de Sitter, and how they may be encoded in the dual CFT.
We would like to return to these questions in the future.

\section{Acknowledgments}

We would like to thank Gabriel Larios, Samir Mathur, Malcolm Perry,
Matt Reece, Dane Rohrer, Irene Valenzuela, Bernard Whiting and 
Haoyu Zhang
for valuable feedback at various stages of this work.
S.C. would like to thank the Harvard University Department of Physics and the Kavli Institute for Theoretical Physics for hospitality and support throughout most of this work.
The work of C.N.P. is supported in part by the DOE Grant No. DE-SC0010813.
The work of M.C. is supported by the DOE (HEP) Award DE-SC0013528, the Simons Foundation Collaboration grant 724069 on ``Special Holonomy in Geometry, Analysis and Physics", the Slovenian Research Agency (ARRS No. P1-0306) and the Fay R. and Eugene L. Langberg Endowed Chair funds.
The work of S.C. was supported in part by the National Science Foundation under Grant No. PHY-2210271. 
This research was supported in part by the National Science Foundation under Grant No. NSF PHY-1748958.

\appendix

\section{Numerical Results}\label{numericalsec}

In this appendix, we collect a few representative plots constructed by
first obtaining numerically-integrated solutions of eqn (\ref{feqn}) 
for the function $f(x)$, for some representative values of the dilaton coupling
$a$.

  Firstly, in Figure 1, we give a plot of $f(x)$ for an example value 
of $a$.  Since the function $f(x)$ looks broadly similar for all
values of $a$ it is not particularly instructive to display it for
different values.  We have chosen to plot it for $a=\ft12$.

\begin{figure}[H]
\begin{center}
\includegraphics[width=300pt]{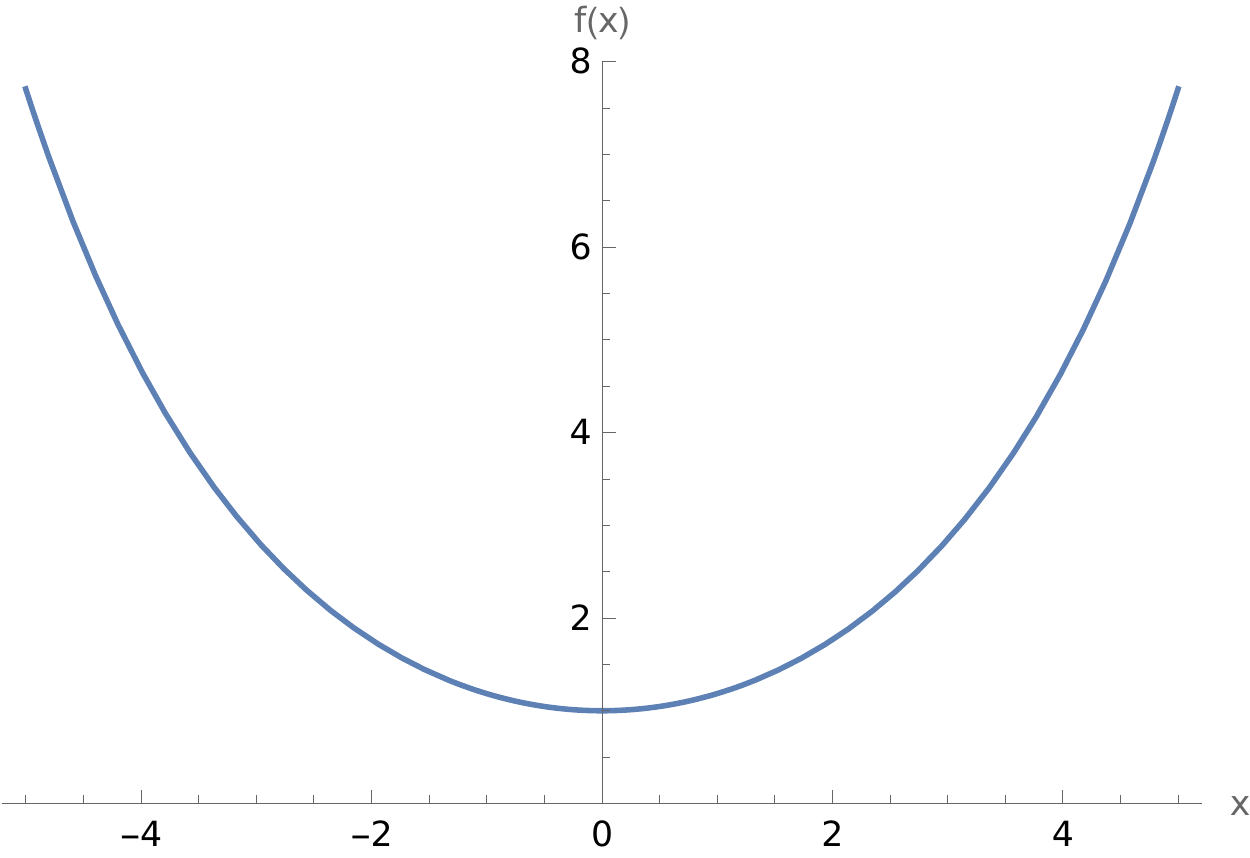}\ \
\end{center}
\caption{\it The function $f(x)$, calculated numerically for $a=\ft12$.}
\end{figure}

  Figure 2 is a plot of the curves $\cF(u,v)=1$ for representative 
choices of the dilaton coupling $a$, where $u=Q/M$ and $v=P/M$.  
The curves were obtained by solving eqn 
(\ref{feqn}) numerically for the various choices of $a$ and then 
plotting the parametric curve $\big(u(x),v(x)\big)$ 
as given by eqn (\ref{uxvx}).

\begin{figure}[H]
\begin{center}
\includegraphics[width=300pt]{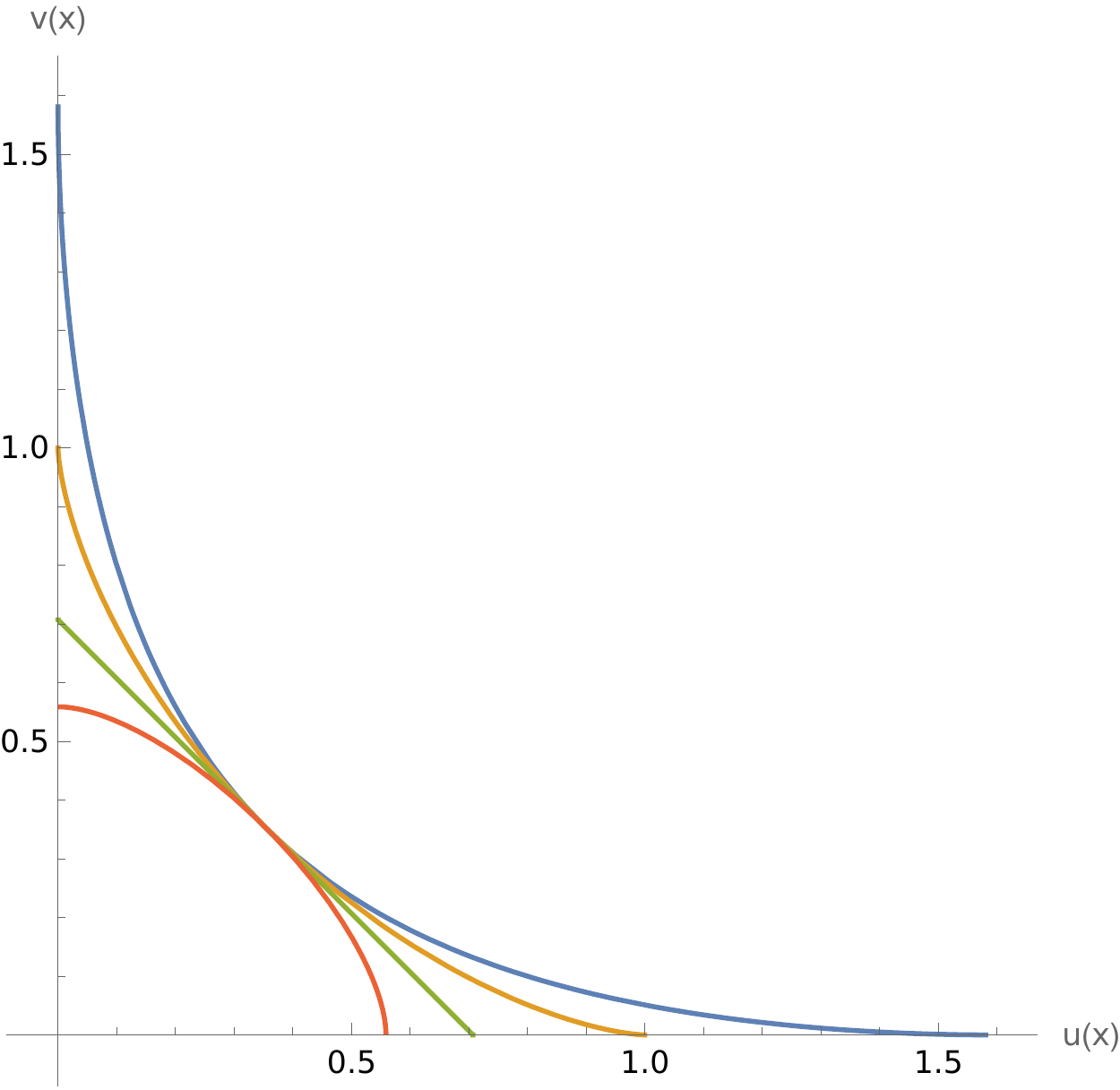}\ \
\end{center}
\caption{\it Plots of $\cF(u,v)=1$ for $a=\ft12$ (convex); 
$a=1$ (flat); $a=\sqrt3$ (concave); and 
$a=3$ (concave, outside the $a=\sqrt3$ curve). Note that the endpoints of the 
curves occur at $u=\ft12\sqrt{1+a^2}$, $v=0$ when $x=\infty$ and
$u=0$, $v=\ft12\sqrt{1+a^2}$ when $x=-\infty$.  All the curves pass through
the point $u=v=\fft1{\sqrt2}$ when $x=0$, which corresponds to the $Q=P$
dyonic Reissner-Nordstr\"om solution for all values of $a$.}
\end{figure}

In Figure 3, we plot the function $a^2\, f(x)-f''(x)$ for a couple of
values of $a$.  The sign of this governs the sign of the radius
of curvature of the curve $\cF(u,v)=1$.

\begin{figure}[H]
\ \ \ \ \ \includegraphics[width=6.5cm]{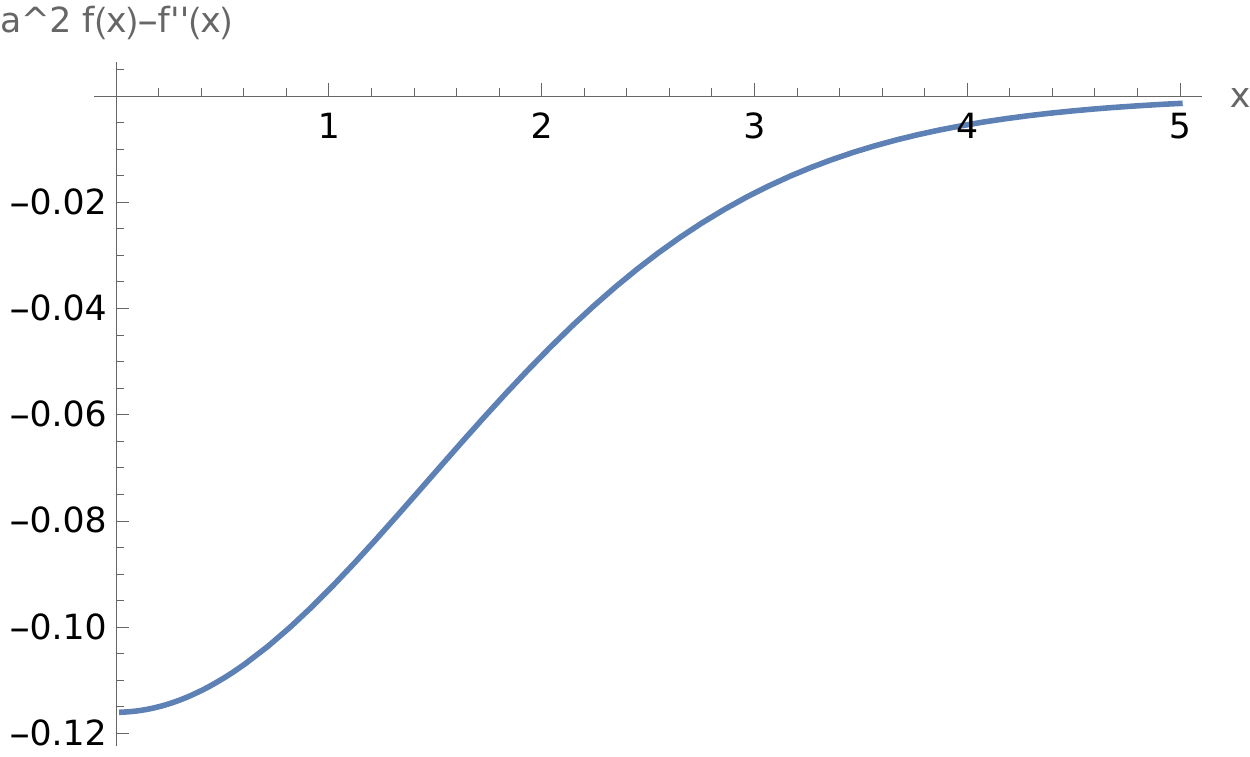}\ \ \ \ \ \ \ \ \
\includegraphics[width=6.5cm]{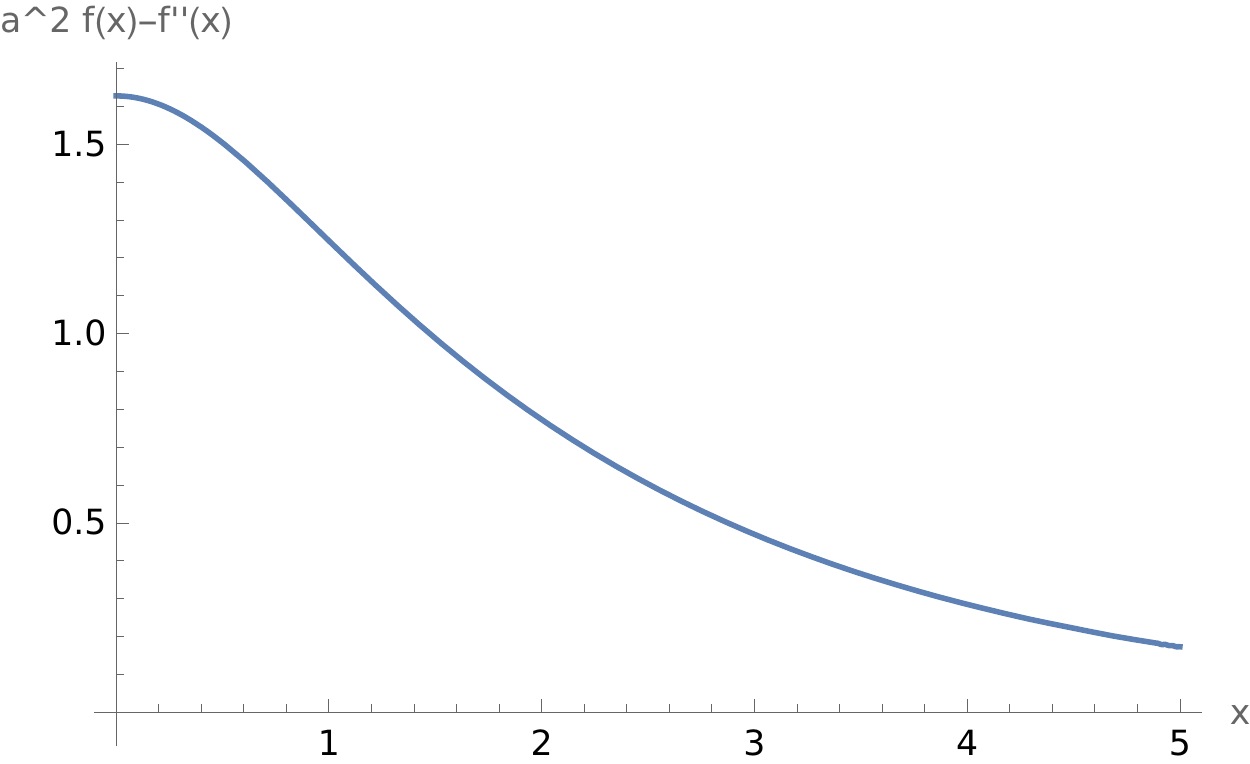}
\caption{\it The function $a^2 f(x)-f''(x)$ plotted for $a=\ft12$ (left-hand 
figure) and $a=2$ (right-hand figure).  These illustrate the fact that
this function is negative for a value of $a<1$, and positive for a value
of $a>1$. From eqn (\ref{curveval}) this shows that the curve is convex for
$a=\ft12$ and concave for $a=2$.}
\label{rmposnfsfs}
\end{figure}

\section{Positivity Proofs for Binding Energy Near $a=1$}\label{DMappendix}

  Here we establish some results for the positivity of the coefficient of
$\ep$ in eqn (\ref{DMneara=1}).  Recall that, as discussed in section
\ref{DMseca1}, the parameter space of the charges $(Q_1,P_1)$ and 
$(Q_2,P_2)$ of the two extremal black holes in this discussion can,
without loss of generality, 
be fully characterised by the three non-trivial parameters $x_1$, 
$\hat x$ and $x_2$ subject to the condition (\ref{xbounds}), i.e. 
$x_2\le \hat x \le x_1$, together with the charge $P_1$, which just
enters the expression for the binding energy as an overall scaling factor.  
(See eqns (\ref{QPxdefs}) for the definitions of
$x_1$, $\hat x$ and $x_2$ in terms of the charges.)

  Taking into account the signs of the 
$\sinh$ denominators in eqn (\ref{DMneara=1}), it can be seen that
the propositions that we wish to
establish are as follows: 

\begin{itemize}

\item[(1)] \underline{If $0\le x_2 \le x_1$} then $W(x_1,\hat x,x_2)\ge0$.

\item[(2)] \underline{If $x_2\le 0\le x_1$} then $W(x_1,\hat x,x_2)\ge0$ if 
$x_2\le \hat x\le 0$ and $W(x_1,\hat x,x_2)\le0$ if $0\le \hat x\le x_1$.

\item[(3)] \underline{If $ x_2\le x_1\le 0$} then $W(x_1,\hat x,x_2)\le0$. 
 
\end{itemize}

  The strategy that we shall follow in order to establish these properties
of the function $W(x_1,\hat x,x_2)$ is to think first of fixed endpoints
$x_1$ and $x_2$, with $x_2\le x_1$, and then allow $\hat x$ to range in the
interval between the endpoints.  An important feature of the function
$W(x_1,\hat x,x_2)$, defined in eqn (\ref{Wdef}), is that although it
depends on $x_1$, $\hat x$ and $x_2$ both as arguments of hyperbolic
functions and with linear dependence as prefactors, it has the property
that after differentiating with respect to any of $x_1$, $\hat x$ or $x_2$,
the resulting function has dependence on that coordinate only as an argument
in hyperbolic functions.  This will mean that we can easily and
explicitly solve for the locations of stationary points of 
$W(x_1,\hat x,x_2)$.

  To establish proposition 1, we first note that $W(x_1,\hat x,x_2)$, defined
in eqn (\ref{Wdef}), vanishes
when $\hat x=x_2$ or $\hat x=x_1$.  Next, viewing $W(x_1,\hat x,x_2)$ as
a function of $\hat x$, we look for the values of $\hat x$ for which
$\fft{d}{d\hat x} W(x_1,\hat x,x_2)=0$.  This gives a quadratic equation
for $\hat d\equiv e^{2\hat x}$:
\bea
&&\hat d^2\, (x_1-x_2 + x_2\,e^{-2 x_1} - x_1\, e^{-2 x_2}) -
   4 \hat d\, \sinh x_1\, \sinh x_2\, \sinh(x_1-x_2) \nn\\
&&+
 x_1-x_2 + x_2\, e^{2 x_1} - x_1\, e^{2 x_2}=0\,.
\eea
It can easily be verified that only one for the two roots for $\hat d$ 
corresponds to a value of $\hat x$ that lies inside the interval
$x_2\le \hat x\le x_1$.  Thus we know that $W(x_1,\hat x,x_2)$, viewed
as a function of $\hat x$, vanishes at $\hat x=x_2$ and $\hat x=x_1$, and
it has just one stationary point in the interval $x_2\le \hat x\le x_1$.
It remains only to establish whether $W(x_1,\hat x,x_2)$ increases from 0 to
a positive maximum and then decreases to 0 again as $\hat x$ ranges from $x_2$
to $x_1$, or whether instead it decreases from 0 to a negative minimum and then
increases to 0 again. This question can be settled by looking at the
sign of $H_2(x_1,x_2)\equiv 
\fft{d}{d\hat x}  W(x_1,\hat x,x_2)\big|_{\hat x=x_2}$.

   Viewing $H_2(x_1,x_2)$ as a function of
$x_1\ge x_2$ with $x_2$ held fixed, its can be seen that $H_2(x_1,x_2)$ 
vanishes when $x_1=x_2$, and that $\fft{d}{dx_1} 
H_2(x_1,x_2)$ has two zeros, when 
\bea
e^{2x_1} = e^{2 x_2}\,,\qquad \hbox{and}\quad e^{2x_1} = \Big[1-
 \fft{2(\sinh2x_1-2x_1)}{e^{2x_2} - 1 - 2 x_2}\Big]\, e^{2x_2}\,.
\eea
The first root lies at $x_1=x_2$, while the second occurs for $x_1<x_2$,
which lies outside the range $x_2\le x_1$ that we are considering. With
$H_2(x_1,x_2)$ and $\fft{d}{dx_1} H_2(x_1, x_2)$ both vanishing at $x_1=x_2$
we turn to the second derivative, finding
\bea
\fft{d^2}{dx_1^2} H_2(x_1,x_2)\Big|_{x_1=x_2}= 4( \sinh 2 x_2- 2 x_2)\,.
\eea
Since $x_2$ is assumed to be non-negative this implies that the
second derivative is non-negative, and therefore since $H_2(x_1,x_2)$ has no turning points for $x_1>x_2$, it follows that we must have 
\bea
H_2(x_1,x_2)\ge 0\qquad \hbox{for}\quad x_1\ge x_2\,.
\eea
Consequently, we have shown that when $0\le x_2\le x_1$, the function
$W(x_1,\hat x, x_2)$ obeys
\bea
W(x_1,\hat x, x_2)>0\qquad \hbox{for}\quad x_2<\hat x<x_1\,,
\eea
with $W(x_1,\hat x, x_2)$ being equal to zero for $\hat x=x_2$ and 
$\hat x=x_1$.  This completes the proof of proposition 1. 
  
  To establish proposition 2, which is for the case where $x_2\le 0\le x_1$,
we note that $W(x_1,\hat x, x_2)$ vanishes when $\hat x=x_2$, when 
$\hat x=x_1$, and when $\hat x=0$.  Since, as we showed previously,
$\fft{d}{d\hat x} W(x_1, \hat x,x_2)$ vanishes at just two values of 
$\hat h$, it must therefore be that one of these roots lies in the
range $x_2<\hat x <0$ and the other in the range $0<\hat x<x_1$.  This
means that one possibility is that 
$W(x_1,\hat x,x_2)$ increases from 0 at $\hat x=x_2$,
then falls to 0 again at $\hat x=0$, and then decreases as $\hat x$ becomes positive, before rising to zero again as $\hat x$ reaches $x_1$.  The
other possibility is that $W(x_1,\hat x,x_2)$ decreases from 0 at $\hat x=x_2$,
then rises to 0 again at $\hat x=0$, and then increases as $\hat x$ becomes     positive, before falling to zero again as $\hat x$ reaches $x_1$.  To settle 
which of these occurs, we can examine the sign of $\fft{d}{d\hat x}
W(x_1,\hat x, x_2)$ at $\hat x=0$. 

  Defining $H_0(x_1,x_2)=\fft{d}{d\hat x} W(x_1,\hat x, x_2)\Big|_{\hat x =0}$,
it can be seen that $\fft{d}{dx_1} H_0(x_1,x_2)=0$ at two values of $x_1$,
namely where
\bea
e^{2x_1}=1\,,\qquad \hbox{and}\quad e^{2x_1} = 
\fft{e^{2 x_2} -1-2x_2}{e^{-2x_2} -1 + 2x_2}\,.
\eea
Writing $x_2$, which by assumption is negative here, as $x_2=-\ft12 p$ where
$p>0$, we see that the second root is at
\bea 
e^{2x_1}= \fft{e^{-p} -1+p}{e^p-1-p}= \fft{\ft12 p^2 - \ft16 p^3 +
  \ft1{24} p^4+\cdots}{\ft12 p^2 + \ft16 p^3 + \ft1{24} p^4+\cdots} \,\, <1\,,
\eea
and therefore this root occurs for $x_1<0$, which is outside the assume range
$x_1>0$.  Thus the function $H_0(x_1,x_2)$, viewed as a function of $x_1$, 
has no turning points in the range $x_1>0$.  

    With $H_0(x_1,x_2)$ and $\fft{d}{dx_1} H_0(x_1,x_2)$ both vanishing at
$x_1=0$, we calculate the second derivative, finding
\bea
\fft{d^2}{dx_1^2} H_0(x_1,x_2)\Big|_{x_1=0} =4(\sinh 2x_2-2x_2)\,,
\eea
which is negative since $x_2$ is assumed to be negative here.  Therefore
since $H_0(x_1,x_2)$ has no turning points when $x_1$ is positive, it
follows that $H_0(x_1,x_2)$ is negative for all $x_1>0$.  Thus we have
shown that the sign of $\fft{d}{d\hat x} W(x_1,\hat x,x_2)$ is negative
when $\hat x=0$.  By the earlier argument, we therefore have that
in this $x_2<0<x_1$ case under discussion,
\bea
W(x_1,\hat x, x_2) >0&& \qquad\hbox{when}\quad x_2<\hat x<0\,,\nn\\
W(x_1,\hat x, x_2) <0&& \qquad\hbox{when}\quad 0<\hat x<x_1\,.
\eea
This completes the proof of proposition 2.

  Finally, to establish proposition 3 we note that the function
$W(x_1,\hat x, x_2)$ has the symmetry $W(-x_2,-\hat x,-x_1)=-
W(x_1,\hat x, x_2)$.  Therefore having already established
in proposition 1 that when $0\le x_2<\hat x \le x_1$ it must be that
$W(x_1,\hat x, x_2)\ge 0$, it immediately follows that when $x_2\le \hat x
\le x_1\le 0$ it must be that $W(x_2,\hat x,x_1)\le 0$.  This proves
proposition 3.

  In summary, the results above provide a general proof that the binding energy 
$\Delta M$ 
defined by eqn (\ref{DeltaM}) is positive when the dilaton coupling is
of the form $a=1+\ep$ and $\ep$ is small and positive, and $\Delta M$ is
negative when $\ep$ is small and negative. 



\begin{thebibliography}{99}

\bibitem{Arkani-Hamed:2006emk}
N.~Arkani-Hamed, L.~Motl, A.~Nicolis and C.~Vafa,
{\it The String landscape, black holes and gravity as the weakest force},
JHEP \textbf{06}, 060 (2007)
doi:10.1088/1126-6708/2007/06/060
[arXiv:hep-th/0601001 [hep-th]].

\bibitem{Harlow:2022gzl}
D.~Harlow, B.~Heidenreich, M.~Reece and T.~Rudelius,
{\it The Weak Gravity Conjecture: A Review},
[arXiv:2201.08380 [hep-th]].


\bibitem{Cremonini:2021upd}
S.~Cremonini, C.R.T.~Jones, J.T.~Liu, B.~McPeak and Y.~Tang,
{\it Repulsive black holes and higher-derivatives},
JHEP \textbf{03}, 013 (2022)
doi:10.1007/JHEP03(2022)013
[arXiv:2110.10178 [hep-th]].

\bibitem{Etheredge:2022rfl}
M.~Etheredge and B.~Heidenreich,
{\it Derivative Corrections to Extremal Black Holes with Moduli},
[arXiv:2211.09823 [hep-th]].

\bibitem{Palti:2017elp}
E.~Palti,
{\it The Weak Gravity Conjecture and Scalar Fields},
JHEP \textbf{08}, 034 (2017)
doi:10.1007/JHEP08(2017)034
[arXiv:1705.04328 [hep-th]].


\bibitem{Heidenreich:2019zkl}
B.~Heidenreich, M.~Reece and T.~Rudelius,
{\it Repulsive Forces and the Weak Gravity Conjecture},
JHEP \textbf{10}, 055 (2019)
doi:10.1007/JHEP10(2019)055
[arXiv:1906.02206 [hep-th]].











\bibitem{Heidenreich:2020upe}
B.~Heidenreich,
{\it Black Holes, Moduli, and Long-Range Forces},
JHEP \textbf{11}, 029 (2020),
doi:10.1007/JHEP11(2020)029,
[arXiv:2006.09378 [hep-th]].


\bibitem{Cremonini:2022sxf}
S.~Cremonini, M.~Cveti\v c, C.N.~Pope and A.~Saha,
{\it Long-range forces between nonidentical black holes with non-BPS 
extremal limits},
Phys. Rev. D \textbf{106}, no.8, 086007 (2022),
doi:10.1103/PhysRevD.106.086007,
[arXiv:2207.00609 [hep-th]].

\bibitem{poltwawil} S.J.~Poletti, J.~Twamley and D.L.~Wiltshire,
{\it Dyonic dilaton black holes},
Class. Quant. Grav. \textbf{12}, 1753-1770 (1995),
[erratum: Class. Quant. Grav. \textbf{12}, 2355 (1995)],
doi:10.1088/0264-9381/12/7/017
[arXiv:hep-th/9502054 [hep-th]].


\bibitem{galkhrorl} D.~Gal'tsov, M.~Khramtsov and D.~Orlov,
{\it ``Triangular'' extremal dilatonic dyons},
Phys. Lett. B \textbf{743}, 87-92 (2015),
doi:10.1016/j.physletb.2015.02.017,
[arXiv:1412.7709 [hep-th]].


\bibitem{gibbmaed} G.W.~Gibbons and K.i.~Maeda,
{\it Black Holes and Membranes in Higher Dimensional Theories with 
Dilaton Fields}, 
Nucl. Phys. B \textbf{298}, 741-775 (1988),
doi:10.1016/0550-3213(88)90006-5.

\bibitem{lupoxu} H.~L\"u, C.N.~Pope and K.W.~Xu,
{\it Liouville and Toda solutions of M theory},
Mod. Phys. Lett. A \textbf{11}, 1785-1796 (1996),
doi:10.1142/S0217732396001776,
[arXiv:hep-th/9604058 [hep-th]].

\bibitem{gibbwilt} G.W.~Gibbons and D.L.~Wiltshire,
{\it Black Holes in Kaluza-Klein Theory},
Annals Phys. \textbf{167}, 201-223 (1986),
[erratum: Annals Phys. \textbf{176}, 393 (1987)],
doi:10.1016/S0003-4916(86)80012-4

\bibitem{gibbkall} G.W.~Gibbons and R.E.~Kallosh,
{\it Topology, entropy and Witten index of dilaton black holes},
Phys. Rev. D \textbf{51}, 2839-2862 (1995),
doi:10.1103/PhysRevD.51.2839,
[arXiv:hep-th/9407118 [hep-th]].

\bibitem{lustea} N.~Cribiori, M.~Dierigl, A.~Gnecchi, D.~Lust and 
M.~Scalisi,
{\it Large and small non-extremal black holes, thermodynamic dualities, 
and the Swampland},
JHEP \textbf{10}, 093 (2022),
doi:10.1007/JHEP10(2022)093,
[arXiv:2202.04657 [hep-th]].


\bibitem{gibkalkol}
G.W.~Gibbons, R.~Kallosh and B.~Kol,
{\it Moduli, scalar charges, and the first law of black hole thermodynamics},
Phys. Rev. Lett. \textbf{77}, 4992-4995 (1996),
doi:10.1103/PhysRevLett.77.4992,
[arXiv:hep-th/9607108 [hep-th]].

\bibitem{giant} W.J.~Geng, B.~Giant, H.~L\"u and C.N.~Pope,
{\it Mass of Dyonic Black Holes and Entropy Super-Additivity},
Class. Quant. Grav. \textbf{36}, no.14, 145003 (2019),
doi:10.1088/1361-6382/ab26e8,
[arXiv:1811.01981 [hep-th]].

\bibitem{rasheed} D.~Rasheed,
{\it The rotating dyonic black holes of Kaluza-Klein theory},
Nucl. Phys. B \textbf{454}, 379-401 (1995),
doi:10.1016/0550-3213(95)00396-A,
[arXiv:hep-th/9505038 [hep-th]].


\bibitem{Aharony:2021mpc}
O.~Aharony and E.~Palti,
{\it Convexity of charged operators in CFTs and the weak gravity conjecture},
Phys. Rev. D \textbf{104}, no.12, 126005 (2021),
doi:10.1103/PhysRevD.104.126005,
[arXiv:2108.04594 [hep-th]].

\bibitem{Andriolo:2022hax}
S.~Andriolo, M.~Michel and E.~Palti,
{\it Self-binding energies in AdS},
JHEP \textbf{02}, 078 (2023),
doi:10.1007/JHEP02(2023)078,
[arXiv:2211.04477 [hep-th]].

\bibitem{Orlando:2023ljh}
D.~Orlando and E.~Palti,
{\it Goldstone Bosons and Convexity},
[arXiv:2303.02178 [hep-th]].


\end{thebibliography}
\end{document}